\documentclass[nofootinbib,twocolumn,superscriptaddress,a4paper]{revtex4-1}
\usepackage{graphicx}
\usepackage{bm,bbm}
\usepackage{xcolor}
\usepackage{tcolorbox}
\usepackage{algorithm}
\usepackage{algpseudocode}
\usepackage{amsmath}
\usepackage{booktabs} 
\usepackage[colorlinks,breaklinks]{hyperref}
\usepackage{lineno}

\makeatletter
\newcommand\org@hypertarget{}
\let\org@hypertarget\hypertarget
\renewcommand\hypertarget[2]{%
\Hy@raisedlink{\org@hypertarget{#1}{}}#2%
  }
\makeatother

\begin{document}

\title{Nonlocal photonic quantum gates over 7.0 km}
\author{Xiao Liu}
\email{These three authors contributed equally to this work.}
\affiliation{CAS Key Laboratory of Quantum Information, University of Science and Technology of China, Hefei 230026, China}
\affiliation{CAS Center for Excellence in Quantum Information and Quantum Physics, University of Science and Technology of China, Hefei 230026, China}
\author{Xiao-Min Hu}
\email{These three authors contributed equally to this work.}
\affiliation{CAS Key Laboratory of Quantum Information, University of Science and Technology of China, Hefei 230026, China}
\affiliation{CAS Center for Excellence in Quantum Information and Quantum Physics, University of Science and Technology of China, Hefei 230026, China}
\affiliation{Hefei National Laboratory, University of Science and Technology of China, Hefei 230088, China}
\author{Tian-Xiang Zhu}
\email{These three authors contributed equally to this work.}
\author{Chao Zhang}
\author{Yi-Xin Xiao}
\author{Jia-Le Miao}
\author{Zhong-Wen Ou}
\author{Pei-Yun Li}
\affiliation{CAS Key Laboratory of Quantum Information, University of Science and Technology of China, Hefei 230026, China}
\affiliation{CAS Center for Excellence in Quantum Information and Quantum Physics, University of Science and Technology of China, Hefei 230026, China}

\author{Bi-Heng Liu}
\email{bhliu@ustc.edu.cn}
\author{Zong-Quan Zhou}
\email{zq\_zhou@ustc.edu.cn}
\author{Chuan-Feng Li}
\email{cfli@ustc.edu.cn}
\author{Guang-Can Guo}
\affiliation{CAS Key Laboratory of Quantum Information, University of Science and Technology of China, Hefei 230026, China}
\affiliation{CAS Center for Excellence in Quantum Information and Quantum Physics, University of Science and Technology of China, Hefei 230026, China}
\affiliation{Hefei National Laboratory, University of Science and Technology of China, Hefei 230088, China}

\begin{abstract}
Quantum networks provide a prospective paradigm to connect separated quantum nodes, which relies on the distribution of long-distance entanglement and active feedforward control of qubits between remote nodes. Such approaches can be utilized to construct nonlocal quantum gates, forming building blocks for distributed quantum computing and other novel quantum applications. However, these gates have only been realized within single nodes or between nodes separated by a few tens of meters, limiting the ability to harness computing resources in large-scale quantum networks. Here, we demonstrate nonlocal photonic quantum gates between two nodes spatially separated by 7.0 km using stationary qubits based on multiplexed quantum memories, flying qubits at telecom wavelengths, and active feedforward control based on field-deployed fibers. Furthermore, we illustrate quantum parallelism by implementing the Deutsch-Jozsa algorithm and the quantum phase estimation algorithm between the two remote nodes. These results represent a proof-of-principle demonstration of quantum gates over metropolitan-scale distances and lay the foundation for the construction of large-scale distributed quantum networks relying on existing fiber channels.
\end{abstract}

\date{\today}
\maketitle

\section*{INTRODUCTION}
The development of large-scale quantum networks \cite{Kimble2008,wehner2018quantum}, comprising interconnected quantum nodes, has garnered significant attention as they provide powerful capabilities beyond the reach of classical counterparts, including global quantum communication \cite{Hensen2015Loophole,Yu2020Entanglement,vanLeent2022Entangling,Krutyanskiy2023Entanglement}, distributed quantum computing \cite{DiVincenzo1995Quantum,Ladd2010QC}, and quantum-enhanced sensing \cite{Komar2014A}. For distributed quantum computing, such network-based approaches can break the technical constraints of single quantum devices, providing an efficient method to scale up the systems \cite{Jiang2007distri,Monroe2014distri,li2024highrate}. Crucially, they can leverage the collective power of multiple remote nodes within expansive quantum networks to solve complex computational tasks efficiently \cite{Cirac1999distri,gottesman1999demonstrating,cuomo2020towards,Oh2023Distributed}, and could enable secure cloud quantum computing with completely classical clients \cite{Huang2017blind}. 

Towards distributed quantum networks, an essential function is the execution of nonlocal quantum gates across network nodes \cite{gottesman1999demonstrating,Eisert2000gate,Bartlett2003gate}, which can be implemented with the help of quantum teleportation, avoiding the direct interaction of two remote qubits. Analogous to quantum state teleportation, which has been demonstrated across diverse physical systems and implementations \cite{Pirandola2015review,Hu2023review}, quantum gate teleportation requires pre-shared entanglement between two separated nodes, local two-qubit gates, and active feedforward control of remote qubits through local operations and classical communications (LOCC). Such remote quantum gates have been realized with photonic qubits without LOCC \cite{huang2004optical1,gap2010optical2}, as well as two species of trapped ions \cite{wang2019trappedions} and two superconducting qubits \cite{Chou2018superconducting}, both within a single device. More recently, based on spin-photon quantum logic, a remote quantum gate has been demonstrated between two separated single atoms linked by a 60-m fiber in the same building \cite{Daiss2021singleatom}. A demonstration of nonlocal quantum gates over longer distances represents a great challenge in the construction of large-scale quantum computing networks.

Here, we demonstrate remote quantum gates based on long-distance distributed photonic entanglement, LOCC enabled with long-lived quantum memories, and local two-qubit operations based on multiple-degree-of-freedom encodings on photons. We characterize the nonlocal controlled-NOT (CNOT) gate by measuring its truth table and the fidelity of four Bell states created from separable states. Furthermore, we use the established nonlocal quantum gates to execute the Deutsch-Jozsa algorithm \cite{deutsch1992rapid} and quantum phase estimation algorithm \cite{kitaev1995quantum}, demonstrating the prototype of distributed quantum computing over metropolitan-scale distances.

\section*{RESULTS}

\begin{figure*}[tbph]
\includegraphics [width=0.8\textwidth]{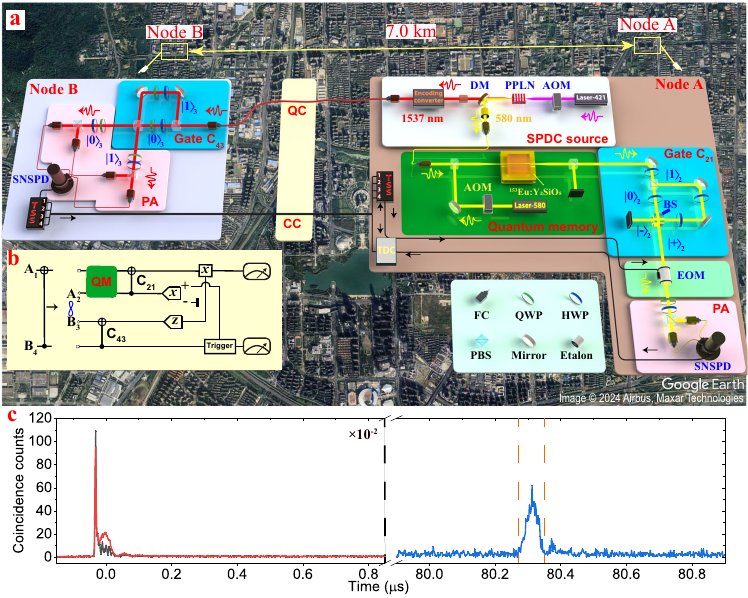}
\caption{Nonlocal quantum gates based on fiber network in the Hefei city. \textbf{a}, Node A is located on the east campus of the University of Science and Technology of China, and node B is located on the foot of the mountain DaShuShan which is 7.0 km away from node A. Pulsed nondegenerate entangled photon pairs are generated through a spontaneous parametric down-conversion (SPDC) process in a periodically-poled lithium niobate (PPLN) waveguide pumped by a 421-nm laser gated with an acousto-optic modulator (AOM). The 580-nm and 1537-nm photons are separated by a dichroic mirror (DM) and spectrally filtered using cascaded etalons. The 580-nm photon is stored in the quantum memory which is implemented using $\mathrm{^{153}Eu^{3+}}$:$\mathrm{Y_2SiO_5}$ crystal cooled by a vibration-isolated cryostat and prepared with modulated laser at 580 nm, while the 1537-nm photon is sent to node B over fiber channel (QC).
The control qubit and target qubit (A$_1$ and B$_4$) are encoded in the polarization degree of freedom (DOF) while the local two-qubit gates (gate $C_{21}$ and $C_{43}$) are implemented between the polarization DOF (A$_1$ and B$_4$) and the path DOF (A$_2$ and B$_3$). The results of polarization analysis (PA) in node B are sent back to node A through another fiber channel (CC) and determine the operations on A$_1$ which is realized using an electro-optic modulator (EOM). FC: fiber collimator, BS: beam splitter, PBS: polarizing beam splitter, HWP: half-wave plate, QWP: quarter-wave plate, TSS: time synchronization system, SNSPD: superconducting nanowire single-photon detector, and TDC: time-to-digital converter. The satellite map is from Google Earth (map data from Airbus and Maxar Technologies).
\textbf{b}, The quantum circuit of nonlocal two-qubit gates between node A and node B. The circuit includes the entanglement between A$_2$ and B$_3$, quantum memories, local two-qubit operations $C_{21}$ and $C_{43}$, measurements of A$_2$ in the $X$ basis and B$_3$ in the $Z$ basis, and classical communication and feedforward operations. 
\textbf{c}, The coincidence counting histograms for the quantum storage of entanglement. 
The red line and the black line represent the transmission of the 580-nm photons, as measured with a 24-MHz transparency window and a 24-MHz AFC memory, respectively. The blue line is the coincidence counts after the 80.315-$\mu$s AFC storage. The data to the left of the black dotted line is multiplied by $10^{-2}$. The retrieved echo peak duration is 72 ns, which is marked as a pair of orange dashed lines. 
}
\label{fig:setup}
\end{figure*}

\textbf{Experimental Setup.} 
The main approach of our experiment is schematically shown in Fig. \ref{fig:setup}. Nondegenerate entangled photon pairsare generated through the spontaneous parametric down-conversion (SPDC) process in node A, which is located on the east campus of the University of Science and Technology of China. The signal photon at 580 nm is stored in a local quantum memory, while the idler photon at 1537 nm is sent along field-deployed fibers to node B, which is located on the foot of the mountain DaShuShan with a spatial separation of 7.0 km from node A. Here, we employ multiple degree of freedoms (DOFs) to encode four qubits with two photons to implement the teleportation-based nonlocal two-qubit gates \cite{gottesman1999demonstrating}. After performing local two-qubit gates at each node and successive LOCC, a nonlocal two-qubit gate is successfully implemented between the distant network nodes. Our protocol involves multiplexed operations in the time domain \cite{Lago-Rivera2021qp,liu2021heralded} and the gate teleportation rate is proportional to the number of stored modes in the quantum memory.

The nondegenerate entangled photon source is based on SPDC in a periodically-poled lithium niobate (PPLN) waveguide pumped by a laser at 421 nm, which is obtained by sum-frequency generation from stabilized and amplified lasers at 580 nm and 1537 nm. Both photons are spectrally filtered with two cascaded etalons to match the bandwidth of the quantum memory (see Section 4 in the Supplementary materials). Energy conservation guarantees that two photons are generated simultaneously, while the precise generation time of the photon pair remains uncertain within the coherence time of the pump laser, resulting in time-energy entanglement \cite{franson1989bell,clausen2011quantum}. After postselection through an unbalanced interferometer of 30 m, the entangled state $\frac{1}{\sqrt{2}}\left(\left|S_{\mathrm{s}} S_{\mathrm{i}}\right\rangle+\left|L_{\mathrm{s}} L_{\mathrm{i}}\right\rangle\right)$ is obtained, where the subscripts s and i represent the signal and idler photons, respectively. If we encode the short $\left|S_{\mathrm{s}, \mathrm{i}}\right\rangle$ and long $\left|L_{\mathrm{s}, \mathrm{i}}\right\rangle$ paths to $|0\rangle$, $|1\rangle$ path encoding, we can obtain the path-entangled state $\frac{1}{\sqrt{2}}(\left|00\right\rangle+\left|11\right\rangle)$. 
Another unbalanced interferometer (encoding converter) is used to convert 1537 nm photons from the time-energy DOF to the polarization DOF before sending them into the field-deployed ultralow-loss optical fiber (see Section 9 in the Supplementary materials). The fiber-optic cable is fixed in a protective underground duct system, maintaining low mechanical vibrations and temperature fluctuations, which enables the long-distance transmission of polarization-encoded photons.

\textbf{Quantum memory and feedforward control.}
The quantum memory is implemented with the atomic frequency comb (AFC) protocol \cite{Afzelius2009Multimode} in a rare-earth-ion-doped crystal (REIC), i.e., 0.2\% doped $\mathrm{^{153}Eu^{3+}}$:$\mathrm{Y_2SiO_5}$ crystal. The isotope $\mathrm{^{153}Eu^{3+}}$ is chosen here to provide a larger storage bandwidth as compared to that of $\mathrm{^{151}Eu^{3+}}$ \cite{Jobez2014Cavity, Ma2021One-hour}. 
The $\mathrm{^{153}Eu^{3+}}$:$\mathrm{Y_2SiO_5}$ crystal is assembled on a close-cycle cryostat with a homemade vibration-isolated sample holder, which allows the preparation of high-resolution AFCs for long-lived and multiplexed photonic storage. The quantum memory should hold the 580-nm photons to wait for the transmission of 1537-nm photons from node A to node B (quantum communication, QC) and the feedback of successive measurement results from node B to node A (classical communication, CC). Both QC and CC are based on field-deployed optical fibers with a length of 7.9 km, which puts a lower bound for the storage time of 79 $\mu$s; therefore, we extend the storage time to 80.315 $\mu$s, which significantly outperforms previous results for photonic entanglement storage (47.7 $\mu$s in Ref. \cite{rakonjac2021entanglement}) in solid-state absorptive quantum memories \cite{clausen2011quantum, Puigibert2020Entanglement, Lago-Rivera2021qp, liu2021heralded, rakonjac2021entanglement, Businger2022Non-classical}. 
Optical memories based on REICs have shown the capability to coherently store light for 1 hour \cite{Longdell2005Stopped,Ma2021One-hour}, making them competitive among various absorptive and emissive quantum memory systems \cite{Lei2023Quantum}.

The prepared AFC has a total bandwidth of $24$ MHz, and the storage efficiency for the bandwidth-matched SPDC source is $(3.2 \pm 0.1) \%$ (Fig. \ref{fig:setup}\textbf{c}). 
Given a single mode duration of 72 ns which covers the retrieved echo peak and the input window length of  79 $\mu$s, the number of temporal modes stored in the memory is 79 $\mu$s / 72 ns = 1097, which results in a linear enhancement of the rate for quantum gate teleportation as compared to the case of employing a single-mode quantum memory  (see Supplementary Fig. 10 in Supplementary materials).

The four qubits employed in the teleportation-based two-qubit gates are denoted as A$_1$, A$_2$ in node A and B$_3$, B$_4$ in node B, where qubits A$_2$ and B$_3$ are prepared in an entangled state in the path DOF of photons, denoted as $|\Phi\rangle_{23}=\frac{1}{\sqrt{2}}\left(|00\rangle_{23}+|11\rangle_{23}\right)$, while the control qubit and target qubit (A$_1$ and B$_4$) are encoded in the polarization DOF of photons. At node A, we use path qubit A$_2$ to perform CNOT operations on polarization qubit A$_1$, followed by measurement of the path qubit A$_2$ along the $X$ basis. At node B, we use polarization qubit B$_4$ to perform CNOT operations on path qubit B$_3$, followed by a measurement of path qubit B$_3$ along the $Z$ basis. Then node B notifies node A of the measurement results through classical communication, and node A performs $I$ or $\sigma_x$ local operations on the polarization qubit A$_1$ using an electro-optic modulator (EOM) to finalize the nonlocal CNOT gate between A$_1$ and B$_4$. In this protocol, without LOCC, the probability of success is $25\%$. The implementation of active feedforward has doubled the success probability of nonlocal two-qubit gates in our experiment. We note that a recent study has also implemented LOCC for multiplexed teleportation of quantum states between a photonic qubit and an atomic memory, linked by 1-km optical fiber, albeit within a single local node \cite{Lago-Rivera2023Long}.
These two-node demonstrations only necessitate quantum memories with preprogrammed storage times. In principle, optical fiber delay loops could provide similar functions with reduced complexity. Nevertheless, quantum memories could provide extended storage times far surpassing those of fiber delay loops and offer versatile multifunctionalities \cite{Lei2023Quantum}, rendering them indispensable components in future large-scale quantum networks.

\begin{figure*}[tbph]
\includegraphics [width=0.9\textwidth]{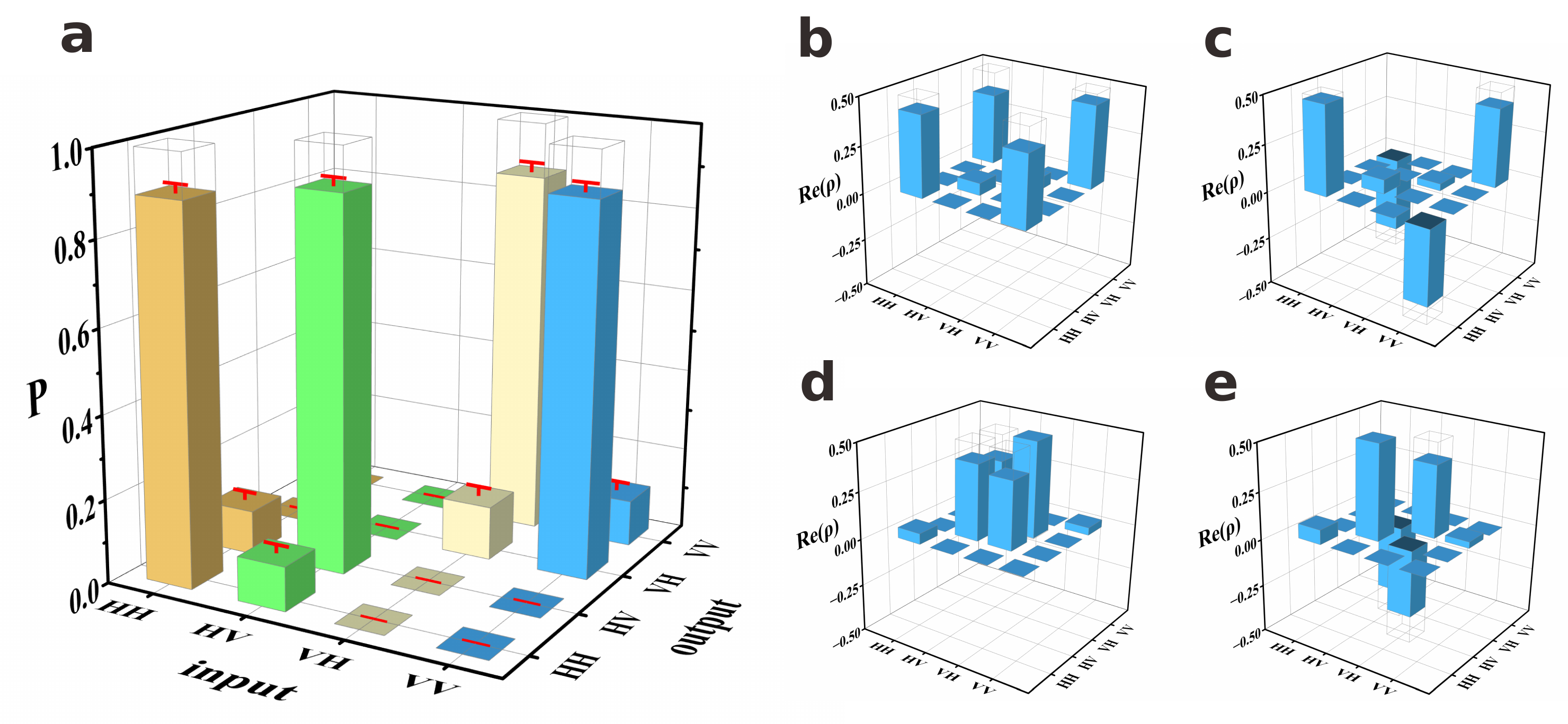}
\caption{Characterization of the nonlocal CNOT gate. \textbf{a}, Truth table of the CNOT gate. The diagram shows the probability P of measuring a certain output state for the four input states: $|HH\rangle$, $|HV\rangle$, $|VH\rangle$, $|VV\rangle$. The expected output states for an ideal CNOT gate are shown as light-shaded bars. \textbf{b}-\textbf{e}, Density matrices of generated Bell states. The diagrams show the real parts of the density matrices of the generated states with inputs of $|+\rangle|H\rangle$, $|+\rangle|V\rangle$, $|-\rangle|H\rangle$, $|-\rangle|V\rangle$, respectively. The ideal Bell states are indicated as shaded bars in the plots. The error bars are one standard deviation.}
\label{fig:truth_table_and_witness_zhengwen}
\end{figure*}

\textbf{Characterization of the nonlocal CNOT gate.}
We use $H$ and $V$ to denote the basis states for polarization qubits A$_1$ and B$_4$. The CNOT gate acts as $|HH\rangle \rightarrow |HH\rangle$, $|HV\rangle \rightarrow |HV\rangle$, and $|VH\rangle \rightarrow |VV\rangle$, $|VV\rangle \rightarrow |VH\rangle$. To characterize this gate, we use all combinations of $|H\rangle/|V\rangle$ for the polarization qubits of node A and node B and then measure the resulting states. 
The truth table of the nonlocal CNOT gate is shown in Fig. \ref{fig:truth_table_and_witness_zhengwen}\textbf{a}, and a fidelity of $(88.7\pm2.1)\%$ compared to the ideal quantum CNOT gate is obtained. 

To demonstrate the quantum properties of the nonlocal CNOT gate, we further use it to create entanglement between two qubits that are initially separable. We initialize qubits A$_1$ and B$_4$ into $|+\rangle|H\rangle$, $|+\rangle|V\rangle$, $|-\rangle|H\rangle$, and $|-\rangle|V\rangle$, respectively, where $|+\rangle=\frac{1}{\sqrt{2}}(|H\rangle+|V\rangle)$, $|-\rangle=\frac{1}{\sqrt{2}}(|H\rangle-|V\rangle)$. An ideal CNOT gate would generate four maximally entangled Bell states $|\Phi^+\rangle$, $|\Phi^-\rangle$, $|\Psi^+\rangle$ and $|\Psi^-\rangle$, where $|\Phi^{\pm}\rangle=|HH\rangle\pm|VV\rangle$ and $|\Psi^{\pm}\rangle=|HV\rangle\pm|VH\rangle$. 
The reconstructed density matrices ($\rho$) of the output states are provided in Fig. \ref{fig:truth_table_and_witness_zhengwen}\textbf{b}-\textbf{e}, with fidelity to expected Bell states of $\mathcal{F}(|\Phi^+\rangle)=(81.1\pm2.6)\%$, $\mathcal{F}(|\Phi^-\rangle)=(85.1\pm2.5)\%$, $\mathcal{F}(|\Psi^+\rangle)=(81.3\pm2.8)\%$ and $\mathcal{F}(|\Psi^-\rangle)=(80.2\pm2.0)\%$, respectively. The average overlap fidelity to the ideal Bell states is $(81.9\pm2.5)\%$, indicating that high-quality entanglement is generated between node A and node B by the nonlocal CNOT gate. The final generation rates of effective nonlocal gates are 0.042 Hz, primarily limited by the efficiency of entangled light sources and quantum memories (see Section 6 in Supplementary materials).

\textbf{Implementation of quantum algorithms.}
Universal quantum computing can be realized through the implementation of the demonstrated nonlocal CNOT gate in conjunction with additional local quantum gates  \cite{gottesman1999demonstrating}. Here, we execute a proof-of-principle for distributed quantum computing using two representative quantum algorithms that exploit quantum advantages: the Deutsch-Jozsa algorithm \cite{deutsch1992rapid} and the phase estimation algorithm \cite{kitaev1995quantum}.

In the Deutsch-Jozsa problem, there are four possible functions $f$ that map one input bit ($a = 0,1$) to one output bit ($f(a) = 0,1$). These functions can be divided into constant functions ($f_{1}(a)=0$, $f_{2}(a)=1$) and balanced functions ($f_{3}(a)=a$, $f_{4}(a)=NOT\ a$). Now, there is an $N$-bit input $x$, and the distinction between the constant and balanced functions $f(x)$ can be achieved through an oracle. Classical algorithms would require querying this oracle $2^{N-1}+1$ times in the worst case for a deterministic answer. In contrast, the Deutsch-Jozsa quantum algorithm requires only a single query in all cases, leveraging inherent quantum parallelism that utilizes quantum interference for computation \cite{deutsch1992rapid,DiVincenzo1995Quantum}, in which the output states might be engineered to be a coherent superposition of states corresponding to different answers.

\begin{figure*}[tbph]
\includegraphics [width=0.9\textwidth]{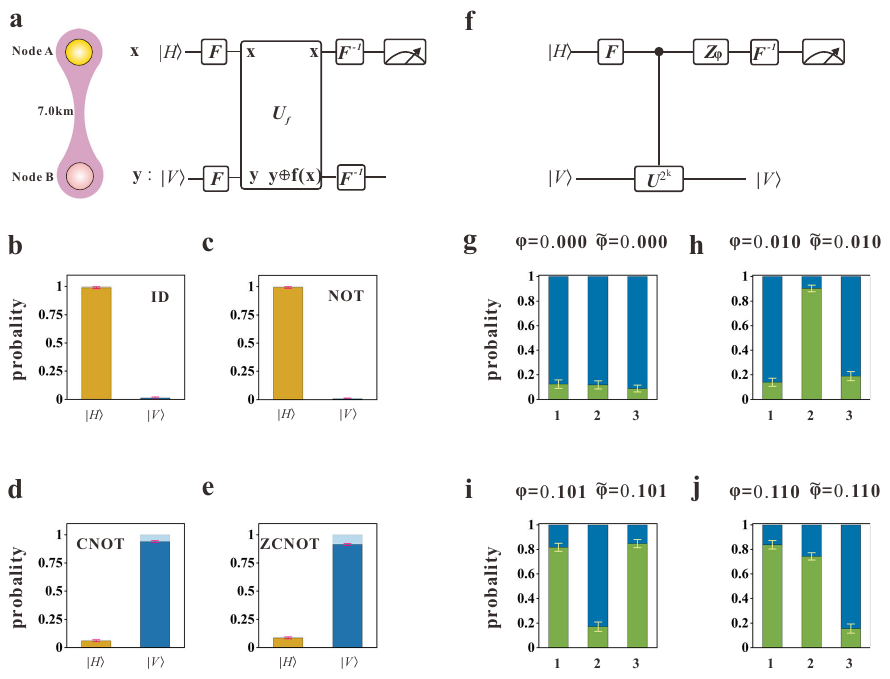}
\caption{
Implementations of the Deutsch-Jozsa algorithm and quantum phase estimation algorithm over two remote network nodes. \textbf{a}, Quantum circuit for implementing the Deutsch–Jozsa algorithm. $x$ is a single query bit, and $y$ is an auxiliary bit, $F$ is the Fourier transform, $F^{-1}$ is the inverse Fourier transform. The box $U_{f}$ represents a unitary operation specific to each of the functions $f$. \textbf{b}, \textbf{c}, The results represent the case of identity (ID) and NOT operations, which belong to the constant function, so the register measurement result collapses to $|H\rangle$. \textbf{d}, \textbf{e}, The results represent the case of the CNOT and zero-CNOT (ZCNOT) operations, which belong to the balance function, so the register measurement result collapses to $|V\rangle$. \textbf{f}, Quantum circuit for the $k$-th iteration of the iterative phase estimation algorithm (IPEA). The algorithm is iterated $m$ times to get an $m$-bit $\tilde{\varphi}$, which is the approximation to the phase of the eigenstate $\varphi$. $Z_{\varphi}$ represents the phase correction as obtained from previous iterations. \textbf{g}-\textbf{j}, Quantum phase estimation results with $m$ = 3.
The corresponding eigenvalues are phase shifts that can be perfectly represented by three binary values with $\varphi$ = 0, $\pi/2$, $5\pi/4$, $3\pi/2$. The blue part represents a phase estimation result of 0, while the green part represents a phase estimation result of 1. The error bars are one standard deviation.}
\label{fig:Algorithm}
\end{figure*}

As shown in Fig. \ref{fig:Algorithm}\textbf{a}, in the two-qubit case \cite{deutsch1985quantum}, the operation that needs to be loaded is $|x\rangle|y\rangle \rightarrow |x\rangle|y \oplus f(x)\rangle$, where $y$ is an auxiliary qubit and $x$ is a single query qubit. Fig. \ref{fig:Algorithm}\textbf{b}-\textbf{e} presents the measured probability distributions of the $x$-register in the computational basis when the function is chosen as constant (Fig. \ref{fig:Algorithm}\textbf{b, c}) and balanced (Fig. \ref{fig:Algorithm}\textbf{d, e}). In all four cases, the experimental fidelity of identifying function classes with one measurement exceeds $91\%$.

The quantum phase estimation algorithm \cite{kitaev1995quantum} is used to estimate the phase of an operator acting on an eigenstate and is frequently used as a subroutine in other quantum algorithms, such as factorization \cite{lanyon2007experimental,politi2009shor} and quantum chemistry \cite{lanyon2010towards}. In this algorithm, the quantum state register consists of a unitary operator $U$ with an eigenstate $|\psi\rangle$ ($U|\psi\rangle=\mathrm{e}^{\mathrm{i}2\pi \varphi}|\psi\rangle$), and information about the unitary operator $U$ is encoded on the measurement register through multiple controlled-$U^{2^{k}}$ operations with $k$ an integer.

The accuracy of the phase estimation algorithm increases with the number of measurement registers. The estimated phase $\tilde{\varphi}$ with $m$ measurement register qubits in binary expansion is $\tilde{\varphi}= 0. \tilde{\varphi}_1 \tilde{\varphi}_2 \ldots \tilde{\varphi}_m$ \cite{kitaev1995quantum}. The circuit with $m$ measurement register qubits can be simplified to an $m$-round iterative phase estimation algorithm (IPEA) \cite{dobvsivcek2007arbitrary} with a single measurement register qubit circuit (Fig. \ref{fig:Algorithm}\textbf{f}). At the end of each iteration, the measurement register qubit is measured, which is an estimate of the $k$-th bit of $\varphi$ in the binary expansion. In the IPEA scheme, the least significant bit is first evaluated (i.e., $k$ iterates backward from $m$ to 1), and then the obtained information is fed back to the phase estimation of subsequent iterations. The iterative information transmission is achieved by rotating $Z_{\varphi}$ of the state register, and its angle is determined by the phase measurement in the previous step. 

The key challenge in implementing a distributed phase estimation algorithm is to achieve nonlocal controlled-\textit{U} (C-\textit{U}) gates. Here, we construct the nonlocal C-\textit{U} gate based on quantum gate teleportation, where a local CNOT gate $C_{21}$ is implemented in node A and $C_{43}$ in node B is changed to a local C-\textit{U} gate (see the Supplementary materials for details). As shown in Fig. \ref{fig:Algorithm}\textbf{g}-\textbf{j}, we perform the quantum phase estimation algorithm for $U=I$, $Z^{1/2}$, $Z^{5/4}$, and $Z^{3/2}$ with three iterations.
These three unitary operations each act on the state $|V\rangle$ with phase shifts $\varphi$ = 0, $\pi/2$, $5\pi/4$, and $3\pi/2$ (corresponding to 0.000, 0.010, 0.101 and 0.110 in binary form, respectively, multiplied by $2\pi$). We obtain the probabilities of each binary digit by polarization measurement of the register. The deviation between measured probabilities and theoretical results is mainly due to the imperfection of nonlocal C-\textit{U} gates. To determine the phase, each binary digit (0 or 1) is chosen based on which measurement probability exceeds 1/2.
Here, three rounds of iteration are sufficient to accurately determine the target phase chosen in this experiment. However, if the precision of the digital phase shift exceeds the number of measurement rounds (e.g., if a phase in binary form 0.1111 is determined but only three measurement rounds are conducted), a discrepancy will arise between the experimental measurements and the actual value.

\section*{DISCUSSION}
To conclude, we have demonstrated nonlocal quantum gates across two nodes separated by 7.0 km and the long-distance active feedforward control of remote
qubits enabled by long-lived quantum memories could serve as a fundamental tool in large-scale quantum networks. 
Several significant improvements are necessary prior to practical applications which include integrating essential functionalities into quantum memories, such as on-demand spin-wave storage \cite{Afzelius2009Multimode,Ma2021Elimination}, enhancing the storage efficiency to outperform optical fiber loops, as well as establishing entangled photon sources and quantum memories at separate nodes to enable remote interconnection via quantum repeaters \cite{Lago-Rivera2021qp, liu2021heralded}. 
It is particularly noteworthy that the incorporation of multiplexed and long-lived quantum memories into quantum networks also opens the opportunity for the investigation of more efficient and high-performance quantum computing systems \cite{Gouzien2021Factoring,liu2023quantum}. 
Due to the use of different DOFs of photons, the photons are measured and absorbed by detectors before feedforward control, and the highest success probability for nonlocal gate operations is 50$\%$, posing a challenge to future scalability. The development of nondestructive photonic qubit detection \cite{Niemietz2021Nondestructive,BrienBrienNondestructive} may offer solutions to these limitations. An alternative approach would be introducing additional photons as communication qubits, allowing target and control qubits retrieved from quantum memories to be available for cascading with more qubits. 
The proof-of-principle demonstration is implemented with photonic qubits, but the principle of linking distant quantum computing nodes with field-deployed fibers can be extended to other platforms such as trapped ions \cite{wang2019trappedions} and neutral atoms \cite{Daiss2021singleatom}, so that to achieve more qubits in a single node and enable deterministic operations. This approach could enable the construction of large-scale quantum computing networks, to make powerful quantum computers that harness the power of quantum communication. 

After the submission of the current work, we note that similar experiments of teleported gates were performed with two silicon T center modules separated by approximately 40 m of fiber \cite{inc2024distributed} and two trapped ion modules separated by approximately 2 m \cite{main2024distributed}.

\section*{METHODS}

\textbf{Quantum memory sample.}
We implement the AFC protocol in an isotope-enriched $\mathrm{^{153}Eu^{3+}}$:$\mathrm{Y_2SiO_5}$ crystal that has a doping concentration of 0.2\% and an isotope $\mathrm{^{153}Eu^{3+}}$ enrichment greater than 99.8\%. The dimensions of the crystal are 17.5 mm $\times$ 5 mm $\times$ 4 mm along the $b$ $\times$ $D1$ $\times$ $D2$ axes. The optical storage utilizes the $\mathrm{{^7}F{_0}\rightarrow{^5}D{_0}}$ transition for site-1 $\mathrm{^{153}Eu^{3+}}$ ions in the $\mathrm{Y_2SiO_5}$ crystal with the background of Earth’s magnetic field. The optical depth of this transition is 5.6, and the inhomogeneous linewidth of the crystal is $(4.6\pm0.1)$ GHz.

\textbf{Witness of entangled states.}
The nonlocal CNOT gate can be used to generate entanglement with separable input states. The real part of the density matrix of the entangled state is measured by a witness \cite{hu2020efficient}. The diagonal terms are given by measuring $\sigma_z^{ij}$, i.e., $\langle i j|\rho| i j\rangle=\langle(|i\rangle\langle i|\otimes|j\rangle\langle j|)\rangle$, where $\rho$ is the density matrix. The off-diagonal terms require a superposition of two subspaces to be measured, i.e., $\Re e[\langle i i|\rho| j j\rangle]=\frac{1}{4}\left(\left\langle\sigma_x^{i j} \otimes \sigma_x^{i j}\right\rangle-\left\langle\sigma_y^{i j} \otimes \sigma_y^{i j}\right\rangle\right)$ and $\Re e[\langle i j|\rho| j i\rangle]=\frac{1}{4}\left(\left\langle\sigma_x^{i j} \otimes \sigma_x^{i j}\right\rangle+\left\langle\sigma_y^{i j} \otimes \sigma_y^{i j}\right\rangle\right)$, where $\sigma_x^{a b}=|a\rangle\langle b|+| b\rangle\langle a|$ and $\sigma_y^{a b}=i|a\rangle\langle b|-i| b\rangle\langle a|$. In the experiment, four Bell states are obtained, and the real parts of some key elements of their density matrix are measured, as shown in Fig. \ref{fig:truth_table_and_witness_zhengwen}\textbf{b}-\textbf{e}. The average fidelity of four Bell states four Bell states is $(81.9\pm2.5)\%$, slightly lower than that of the initial entangled states. The reduction in fidelity is mainly attributed to the low storage efficiency of the quantum memory. The transmission through external optical fibers and the feedforward operation of LOCC also cause a slight decrease in fidelity.
\bigskip

\textbf{Data availability}
Data that support the findings of this study are available from the corresponding authors upon request.
\\

\textbf{Code availability}
The custom codes used to produce the results presented in this paper are available from the corresponding authors upon request.

\bigskip
\textbf{Acknowledgments}
This work is supported by the National Key R\&D Program of China (No. 2017YFA0304100), Innovation Program for Quantum Science and Technology (No. 2021ZD0301200), the National Natural Science Foundation of China (Nos. 12374338, 12222411, 11904357, 12174367, 12204458, 11821404, 12204459 and 62322513), Anhui Provincial Natural Science Foundation (Nos. 2108085QA26 and 2408085JX002), Fundamental Research Funds for the Central Universities, Xiaomi Young Talents Program, Anhui Province science and technology innovation project (No. 202423r06050004), the China Postdoctoral Science Foundation (2023M743400). Z.-Q.Z acknowledges the support from the Youth Innovation Promotion Association CAS. T.-X. Z. acknowledges the support from the Postdoctoral Fellowship Program of CPSF. The allocation of node B and the deployment of ultralow-loss fiber is supported by China Unicom (Anhui).\\

\textbf{Author contributions}
Z.-Q.Z., B.-H.L., and C.-F.L. designed the experiment; X.L. constructed the fiber channel, T.-X. Z. constructed the quantum memory with the help of Y.-X.X. and Z.-W.O. X.-M.H. constructed the quantum light source with the help of C.Z., X.L., J.-L.M. and P.-Y.L. X.-M.H., T.-X.Z., X.L., C. Z., B.-H.L., and Z.-Q.Z. wrote the manuscript with input from others. Z.-Q.Z., B.-H.L., C.-F.L., and G.-C.G. supervised the project. All authors discussed the experimental procedures and results.\\

\textbf{Competing interests}
The authors declare no competing interests.

\end{document}


\title{Supplementary Materials for\\
``Nonlocal photonic quantum gates over 7.0 km"}

\author{Xiao Liu}
\email{These three authors contributed equally to this work.}
\affiliation{CAS Key Laboratory of Quantum Information, University of Science and Technology of China, Hefei 230026, China}
\affiliation{CAS Center for Excellence in Quantum Information and Quantum Physics, University of Science and Technology of China, Hefei 230026, China}
\author{Xiao-Min Hu}
\email{These three authors contributed equally to this work.}
\affiliation{CAS Key Laboratory of Quantum Information, University of Science and Technology of China, Hefei 230026, China}
\affiliation{CAS Center for Excellence in Quantum Information and Quantum Physics, University of Science and Technology of China, Hefei 230026, China}
\affiliation{Hefei National Laboratory, University of Science and Technology of China, Hefei 230088, China}
\author{Tian-Xiang Zhu}
\email{These three authors contributed equally to this work.}
\author{Chao Zhang}
\author{Yi-Xin Xiao}
\author{Jia-Le Miao}
\author{Zhong-Wen Ou}
\author{Pei-Yun Li}
\affiliation{CAS Key Laboratory of Quantum Information, University of Science and Technology of China, Hefei 230026, China}
\affiliation{CAS Center for Excellence in Quantum Information and Quantum Physics, University of Science and Technology of China, Hefei 230026, China}

\author{Bi-Heng Liu}
\email{bhliu@ustc.edu.cn}
\author{Zong-Quan Zhou}
\email{zq\_zhou@ustc.edu.cn}
\author{Chuan-Feng Li}
\email{cfli@ustc.edu.cn}
\author{Guang-Can Guo}
\affiliation{CAS Key Laboratory of Quantum Information, University of Science and Technology of China, Hefei 230026, China}
\affiliation{CAS Center for Excellence in Quantum Information and Quantum Physics, University of Science and Technology of China, Hefei 230026, China}
\affiliation{Hefei National Laboratory, University of Science and Technology of China, Hefei 230088, China}

\begin{abstract}
\bigskip\bigskip\bigskip
\tableofcontents

\end{abstract}
\date{\today}
\maketitle

\subsection{1. Quantum gate teleportation protocol}

We briefly explain the protocol for the teleportation of quantum gates. A schematic diagram of a typical quantum gate teleportation is shown in Supplementary Figure \ref{fig:CNOT GATE}\textbf{a}. Assume that two parties, Alice and Bob, each have two qubits 1, 2, and 3, 4. EPR entanglement $|\Phi\rangle_{23}=\frac{1}{\sqrt{2}}\left(|00\rangle_{23}+|11\rangle_{23}\right)$ is shared between qubits 2 and 3. The purpose is to implement nonlocal CNOT operations for qubit 4 on qubit 1 with the assistance of this EPR entanglement, as well as local two-qubit operations and local single-qubit operations under classical communications. A quantum CNOT gate $C_{43}$ is performed on local qubits 3, 4, and a quantum CNOT gate $C_{21}$ is performed on local qubits 1, 2. Then, qubits 2 and 3 are measured on the $\{+,-\}$ and $\{0, 1\}$ bases, respectively. 
Finally, Alice and Bob will notify each other of the measurement results and perform the required operations on qubits 1 and 4 depending on the results. Quantum gate teleportation is realized by the following convention:

\begin{equation}
\begin{aligned}
C_{43} C_{21}\left(|\Psi\rangle_{14} \otimes|\Phi\rangle_{23}\right) & =|+0\rangle_{23} \otimes C_{41}\left(|\Psi\rangle_{14}\right) \\
& +|+1\rangle_{23} \otimes \sigma_1^x C_{41}\left(|\Psi\rangle_{14}\right) \\
& +|-0\rangle_{23} \otimes \sigma_4^z C_{41}\left(|\Psi\rangle_{14}\right) \\
& +|-1\rangle_{23} \otimes\left(-\sigma_1^x \sigma_4^z\right) C_{41}\left(|\Psi\rangle_{14}\right).
\end{aligned}
\end{equation}

\begin{figure}[tbph]
\includegraphics [width=0.6\textwidth]{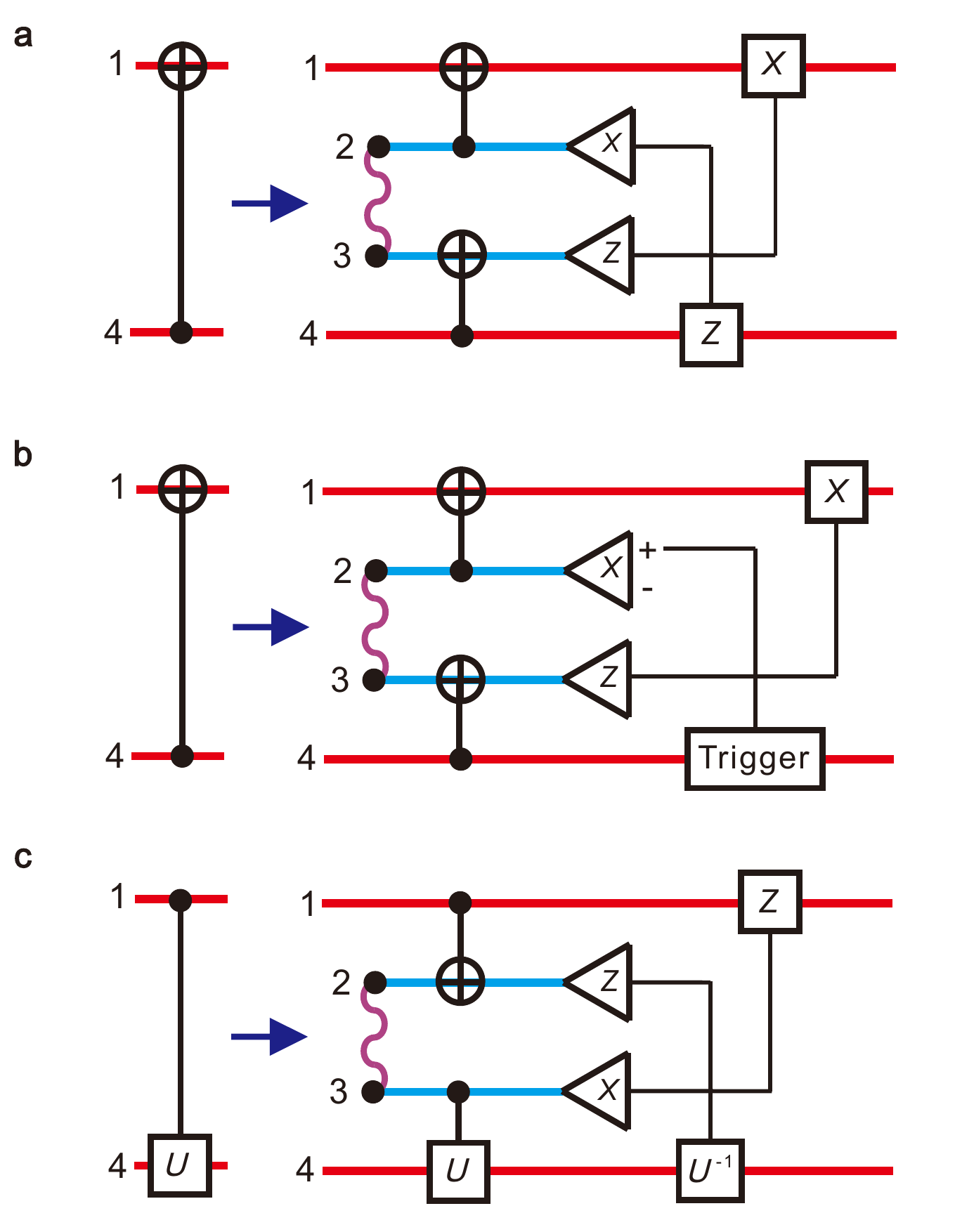}
\vspace{-0.3cm}
\caption{Circuit for quantum teleportation of controlled gates. \textbf{a}, Quantum teleportation of CNOT gate. \textbf{b}, Quantum teleportation of the CNOT gate with unilateral feedforward control. \textbf{c}, Quantum teleportation of the C-\textit{U} gate.}
\label{fig:CNOT GATE}
\vspace{0.2cm}
\end{figure}

Here, we employ two degrees of freedom (DOFs) of two photons to encode the four qubits. 
The local two-qubit gates between different DOFs are inherently deterministic with only linear elements \cite{Kwiat1998Embedded}.
However, when we obtain the measurement results of qubit 2 with single-photon detectors, qubit 1 also collapses. Therefore, as shown in Supplementary Figure \ref{fig:CNOT GATE}\textbf{b}, we discard half of the measurement results of qubit 2, and the quantum gate teleportation is performed with a success probability of $50\%$. Nondestructive measurements of photonic qubits \cite{Niemietz2021Nondestructive,BrienBrienNondestructive} would be required to achieve bidirectional classical feedforward control, and thus to achieve deterministic implementations of this scheme with photons unabsorbed for further operations. 
A different strategy involves introducing additional photons as communication qubits for local two-qubit gates, enabling the control and target qubits for nonlocal quantum gates undetected and available for cascading. In this case, photon number resolving quantum nondemolition detection \cite{Nemoto2004Nearly} or other efficient nonlinear processes \cite{Chang2014Quantum} at the single-photon level are required to obtain higher success probabilities. To scale up the number of qubits and operations at each node, many other accessible DOFs including high-dimensional DOFs could be exploited for interaction \cite{Ren2013Deterministic,Imany2019High}, although practical scalability remains a great challenge \cite{Kwiat1997Hyper,DENG2017Quantum}. A more general approach is to use the interference of different photons to cascade qubits towards universal quantum computing \cite{Sleator1995Realizable} with the Knill-Laflamme-Milburn \cite{Knill2001scheme} scheme.

The realization of quantum teleportation for arbitrary controlled unitary (C-\textit{U}) gates is slightly different from that of quantum teleportation for CNOT gates.
As shown in Supplementary Figure \ref{fig:CNOT GATE}\textbf{c}, qubit 1 performs a CNOT operation on qubit 2, and qubit 3 performs a C-\textit{U} operation on qubit 4. In this way, we implement the nonlocal C-\textit{U} operation applied by qubit 1 to qubit 4. Similarly, we have

\begin{equation}
\begin{aligned}
CU_{34} C_{12}\left(|\Psi\rangle_{14} \otimes|\Phi\rangle_{23}\right)& =|0+\rangle_{23} \otimes CU_{14}\left(|\Psi\rangle_{14}\right) \\
& +|0-\rangle_{23} \otimes \sigma_1^z CU_{14}\left(|\Psi\rangle_{14}\right) \\
& +|1+\rangle_{23} \otimes U^{-1}_4 CU_{14}\left(|\Psi\rangle_{14}\right) \\
& +|1-\rangle_{23} \otimes\left(-\sigma_1^z U^{-1}_4\right) C_{14}\left(|\Psi\rangle_{14}\right),
\end{aligned}
\label{CU}
\end{equation}
with $UU^{-1}=I$. Since photonic qubits are measured with absorptive detectors, here the last two terms of Eq. \ref{CU} are dropped, and the quantum teleportation of the C-\textit{U} gate is also realized with a $50\%$ success probability.

\subsection{2. Overall experimental setup}

In this section, we briefly introduce the entire experimental setup at node A and node B (as shown in Supplementary Figure \ref{fig:scheme sm}). In Supplementary Figure \ref{fig:scheme sm}\textbf{a}, a two-photon source is generated using the spontaneous parametric down-conversion (SPDC) process, detailed in section 4 of the supplementary materials. Supplementary Figure \ref{fig:scheme sm}\textbf{b}, \textbf{d} depict two unbalanced interferometers with arm differences of 30 m, utilized for postselection of entangled states. Additionally, the interferometer in Supplementary Figure \ref{fig:scheme sm}\textbf{b} also serves as an encoding converter between path and polarization DOFs, detailed in section 9 of the supplementary materials, while the interferometer in Supplementary Figure \ref{fig:scheme sm}\textbf{d} implements the CNOT operation between A$_1$ and A$_2$, as explained in section 9 of the supplementary materials. In Supplementary Figure \ref{fig:scheme sm}\textbf{c}, a solid-state quantum memory (QM) with storage time up to 80.315 $\mu$s is employed to store 580-nm photons and await the feedforward signal from the measurement results of node B, thereby achieving an improvement in the success probability of controlled gates (see details in the main text and section 1 of the supplementary materials). The 1537-nm photons in polarization DOF are distributed through a 7.9-km field-deployed fiber to reach node B. In Supplementary Figure \ref{fig:scheme sm}\textbf{e}, the controlled gate of B$_3$ to B$_4$ in node B is implemented, as detailed in section 9 of the supplementary materials. Subsequently, the real-time measurement results of node B are fed back to node A through the classical channel. Based on the measurement results of node B, node A performs the corresponding unitary transformation ($I$ or $\sigma_x$) by a high-speed electro-optic modulator (EOM). Finally, a polarization analyzer is used to analyze the input and output quantum state results.

\begin{figure}[tbph]
\includegraphics [width=1\textwidth]{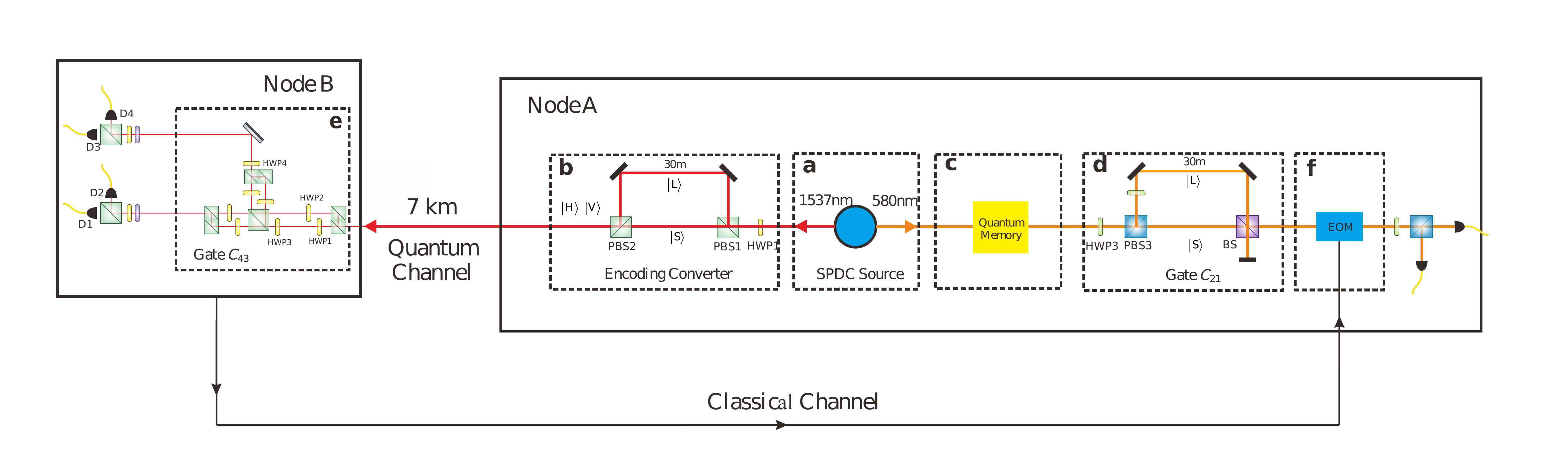}
\vspace{-0.3cm}
\caption{Complete experimental setup. \textbf{a}, SPDC source. \textbf{b}, Ecoding Converter. \textbf{c}, Quantum memory. \textbf{d}, Controlled operation for A$_1$ and A$_2$. \textbf{e}, Controlled operation for B$_3$ and B$_4$. \textbf{f}, Ultra-fast feedforward control on electro-optic modulator (EOM).}
\label{fig:scheme sm}
\vspace{0.2cm}
\end{figure}

\subsection{3. Laser system}

\begin{figure}[tbph]
  \includegraphics [width=0.5\textwidth]{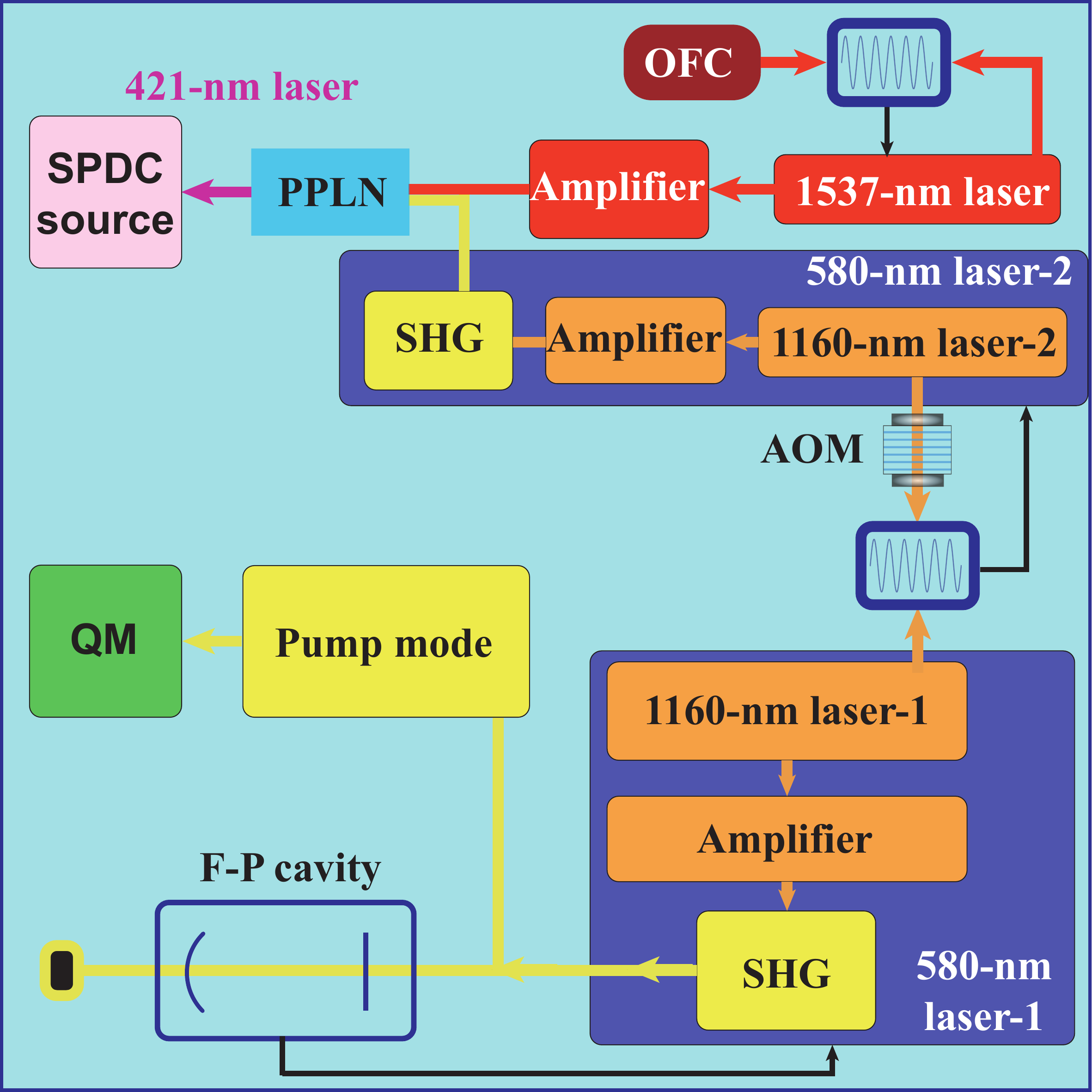}
    \caption{Schematic diagram of the laser system. Two 580-nm lasers are frequency doubled from the 1160-nm seed lasers. The 580-nm laser-1 is employed in the preparation of quantum memory, which is frequency-locked to a temperature-controlled Fabry-Pérot (F-P) cavity placed inside a vacuum chamber. The 421-nm laser serves as the pump laser for the SPDC source, which is obtained by sum-frequency generation in a PPLN crystal with the pump source from the 580-nm laser-2 and a 1537-nm laser. The 580-nm laser-2 is frequency-locked with the 580-nm laser-1 by phase-locking of their seed lasers. The acoustic optical modulator (AOM) adjusts the frequency offset to achieve frequency alignment of the entangled 580-nm photons and the quantum memory (QM). The 1537-nm laser is phase-locked to an optical frequency comb (OFC).
    }
    \label{fig: laser system}
\end{figure}

The main laser system used in the experiment is placed at node A, with a schematic diagram shown in Supplementary Figure \ref{fig: laser system}. Two independent 580-nm lasers, both frequency-doubled from 1160-nm lasers, are employed for the SPDC source and the QM, respectively. The 580-nm laser-1 (Toptica, TA-SHG) used for quantum memory is frequency-locked to a high-finesse Fabry-Pérot (F-P) cavity, achieving a linewidth of approximately 0.3 kHz. The F-P cavity is temperature-controlled within a stability of 0.01 K and is placed in a vacuum chamber maintained by a sputter ion pump. The other 580-nm laser system (580-nm laser-2) consists of an 1160-nm narrow linewidth semiconductor laser (Toptica, DL PRO), a Raman fiber amplifier, and a single-pass frequency doubling system made by another company (Precilasers), with an output power up to 3 W. The 580-nm laser-2 is frequency stabilized by phase locking its 1160-nm seed laser to the 1160-nm seed laser in the 580-nm laser-1. The beat frequency between two 1160-nm lasers can be adjusted by changing the frequency of 1160-nm laser-2 with an acoustic optical modulator (AOM).
A 1537-nm laser is combined with the 580-nm laser-2 in a PPLN bulk crystal, producing a 421-nm laser by sum-frequency generation. The 1537-nm seed laser (Precilasers, EFL-SF-1536-S) is phase-locked to a commercial optical frequency comb (Menlo Systems, FC1500-250-ULN) and amplified by an erbium-doped fiber amplifier, achieving a linewidth of approximately 20 kHz and an output power of 5 W. The 421-nm laser, with a total output power of 200 mW, is employed as the pump laser for the SPDC process.

\subsection{4. SPDC photon source}

The pump light at 421 nm
is injected into a PPLN waveguide, leading to a nondegenerate SPDC process that produces signal photons at 580 nm and idler photons at 1537 nm. The peak pump power before the waveguide is set to approximately 10 mW and is gated into pulses according to the time sequence shown in Supplementary Figure \ref{fig:Time sequence}\textbf{b}. The 580-nm photon is reflected by a dichroic mirror (Semrock, FF775-Di01-25x36) and is filtered by a long-pass filter (Semrock, BLP01-442R-2f) and a band-pass filter centered at 580 nm with a 3-nm bandwidth. The 1537-nm photon passes through the dichroic mirror and is filtered using a filter centered at 1550 nm with a 30-nm bandwidth. We have significantly improved the collection efficiency of entangled photon sources using a 4-$f$ system comprising pairs of cylindrical lenses. Without the use of the cylindrical lenses (that is, removing parts CL@580 nm and CL@1537 nm in Supplementary Figure \ref{fig:source}), the heralding efficiency before the filtering system is $19.5\%$. After the filtering system, the heralding efficiency of the narrow-band source becomes $4.1\%$. To optimize the coupling mode, two pairs of cylindrical lenses are used to amplify the short axis of the light spot for 1537 nm and 580 nm, respectively. After shaping the light, the heralding efficiency of the source before the filtering system is increased to $27.5\%$, while the efficiency of the narrow-band source after the filtering system is $4.9\%$. This represents an overall improvement of approximately $20\%$ by employing the cylindrical lenses. 

\begin{figure}[tbph]
\includegraphics [width=1\textwidth]{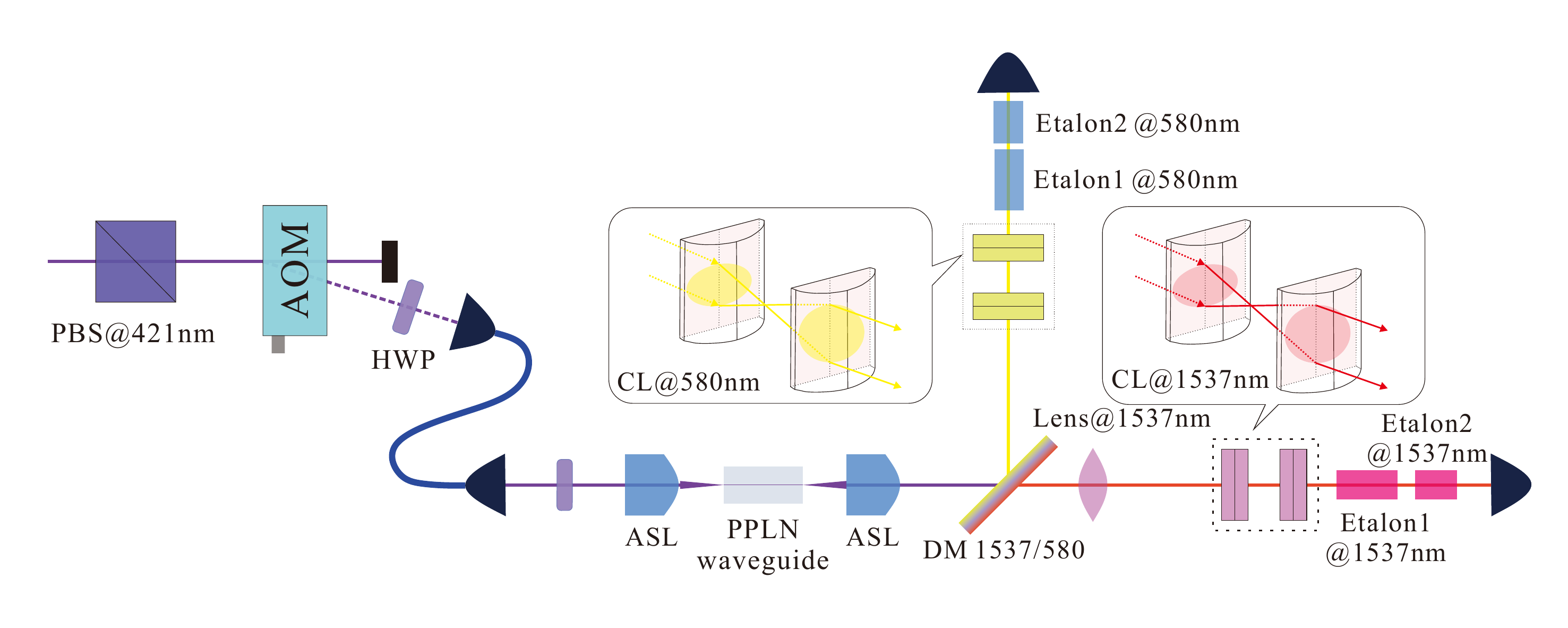}
\caption{Setup of the spontaneous parametric down-conversion (SPDC) source. To reduce random coincidence noise during the storage process, the pump light is gated using an AOM. The pump light at 421 nm is adjusted to horizontal polarization using a half-wave plate (HWP) and focused by an aspheric lens (ASL) with a focal length of $f_1 = 8$ mm before coupling into a PPLN waveguide. In the PPLN waveguide, the nondegenerate SPDC process produces a signal photon at 580 nm and an idler photon at 1537 nm, which are collimated by an ASL with a focal length of $f_2 = 6.24$ mm and then separated by a dichroic mirror (DM). Since the intrinsic mode of light exiting from the waveguide is elliptical, we use cylindrical lenses (CL@580 nm and CL@1537 nm) to shape the short axis of the light spot, aligning the propagation and coupling modes to enhance the collection efficiency. Narrow-band filtering is performed on these photons using cascaded etalons. Etalon1 @580 nm has a bandwidth of 3 GHz and a free spectral range (FSR) of 100 GHz, and etalon2 @580 nm has a bandwidth of 150 MHz and an FSR of 2 GHz. Etalon1 @1537 nm has a bandwidth of 3 GHz and an FSR of 100 GHz, and etalon2 @1537 nm has a bandwidth of 150 MHz and an FSR of 2 GHz. PBS: polarizing beamsplitter.}
\label{fig:source}
\end{figure}

\begin{figure}[tbph]
\includegraphics [width=0.6\textwidth]{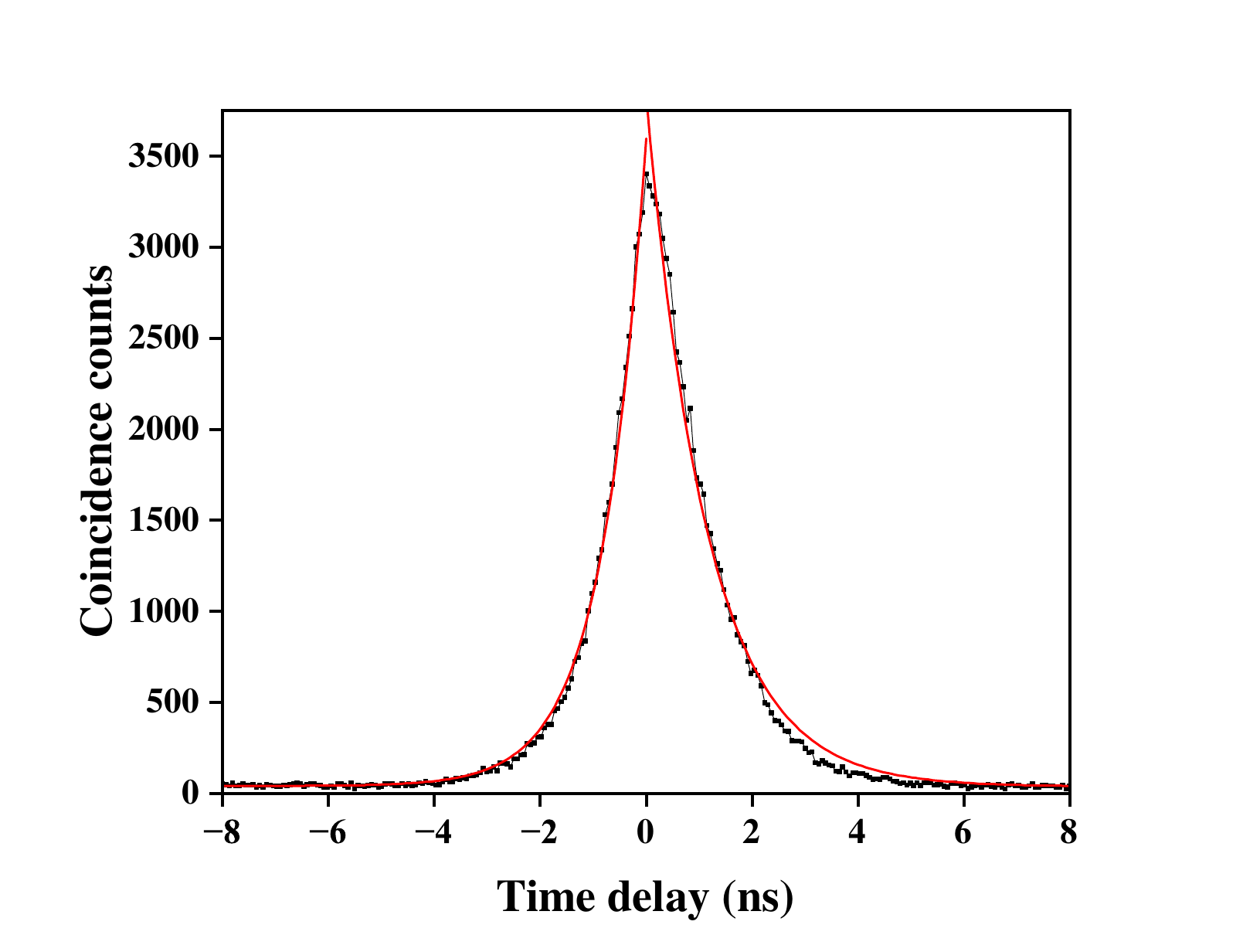}
\caption{Two-photon coincidence counting histogram for the narrow-band entangled photons. The temporal resolution is 64 ps and the integration time is 600 s. The black dot is the raw data, and the red line is the fitted curve using $e^{-2 \pi \Delta v \tau}$.}
\label{fig:g2}
\end{figure}

The quantum correlation between two photons can be expressed as the second-order cross-correlation function $g_{12}^{(2)}(\Delta\tau)=P_{12} /\left(P_1 P_2\right)$ with zero time difference, where $P_{12}$ represents the two-photon coincidence probability in the time window $\Delta\tau$ and $P_{1} (P_{2})$ corresponds to the measured probability of a single photon of port 1 (2). In practice, the correlation between two photons can be obtained by measuring the two-photon coincidence counts, as shown in Supplementary Figure \ref{fig:g2}.
By Lorentz fitting of both sides of the peak value of the correlation function with $e^{-2 \pi \Delta v \tau}$, we estimate photon bandwidths $\Delta v$ of 178 MHz at 580 nm and 155 MHz at 1537 nm. The $g_{12}^{(2)}(\Delta\tau)$ between photon pairs at zero time difference is 36.9 in a detection window of $\Delta\tau = 6.4$ ns. We use a 30-m unbalanced interferometer to generate time-energy entanglement, and observe a visibility of $0.990\pm0.001$ for the computational basis {$|0\rangle$, $|1\rangle$} and $0.862\pm0.001$ for the Fourier basis {$|0\rangle+|1\rangle$,$|0\rangle-|1\rangle$}. This indicates that the fidelity of our initial entangled state is approximately $0.926\pm0.001$.

\subsection{5. Quantum memory}

\begin{figure}[tbph]
	\includegraphics [width=0.9\textwidth]{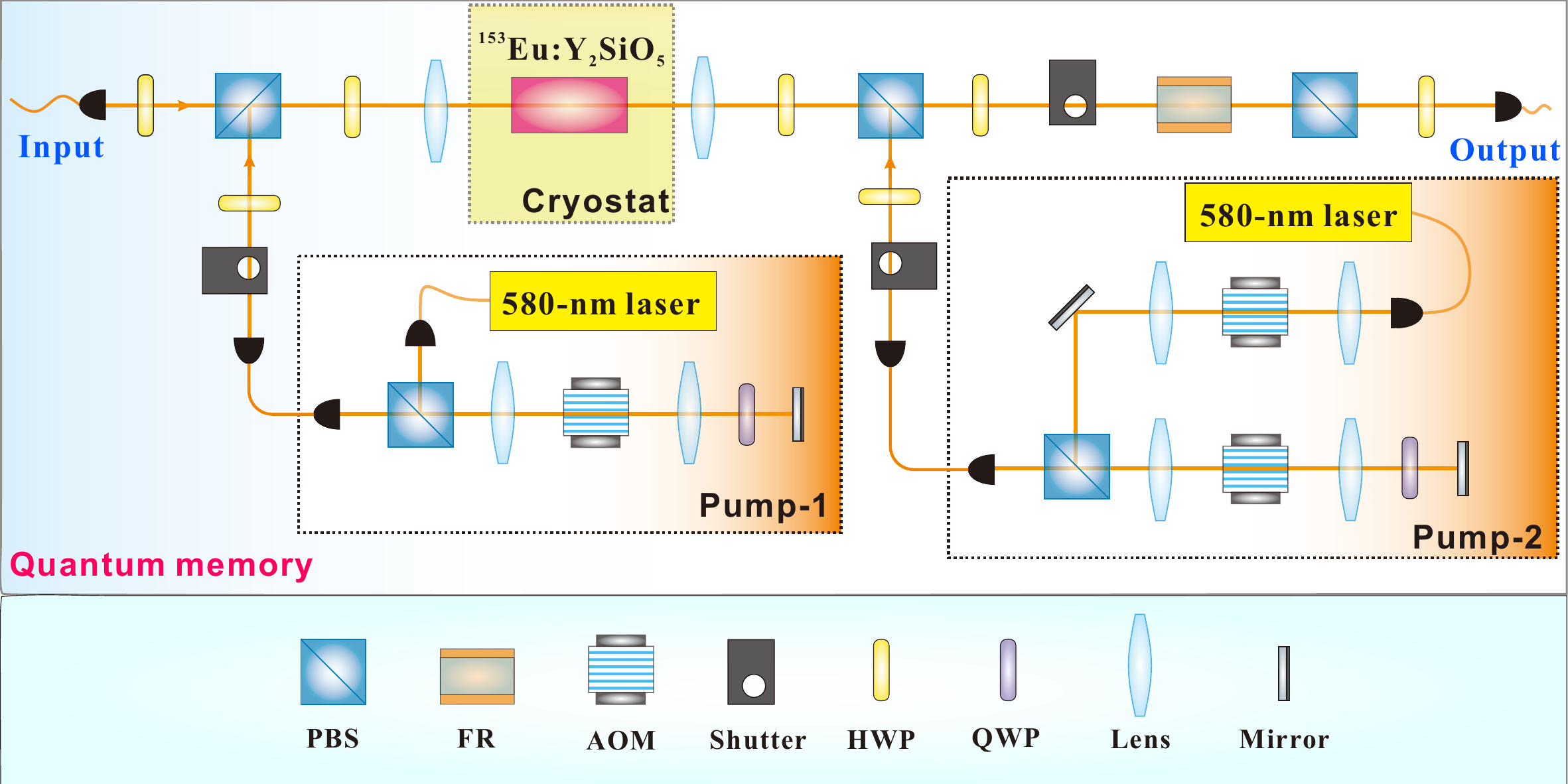}
	\caption{Schematic diagram of the quantum memory. The input mode is polarized along the $D1$ axis of the $\mathrm{^{153}Eu^{3+}}$:$\mathrm{Y_2SiO_5}$ crystal. The crystal is cooled to 3.1 K by a vibration-isolated closed-cycle cryostat. Two counterpropagating pump modes are employed to prepare the absorption profile of the quantum memory. Both pump lights are generated by a stabilized 580-nm laser and modulated by AOMs. The pump modes are polarized along the $D2$ axis of the crystal and coupled into the crystal by a PBS. The Faraday rotator (FR) is employed as an optical isolator, together with mechanical shutters, to protect the superconducting nanowire single-photon detector (SNSPD) from the classical pump light. HWP: half-wave plate; QWP: quarter-wave plate.}
	\label{fig:quantum memory}
\end{figure}

Among various quantum storage protocols \cite{Fleischhauer2000Dark-State, Moiseev2001Complete, Duan2001Long-distance, Afzelius2009Multimode, Ma2021Elimination},
the atomic frequency comb (AFC) \cite{Afzelius2009Multimode} scheme has the particular advantages of large storage bandwidth and high multimode capacity, which is crucial for efficient interface with quantum light sources. The $\mathrm{Eu^{3+}}$:$\mathrm{Y_2SiO_5}$ crystal has been an important candidate for quantum memory because of its extremely long optical and spin coherence lifetimes \cite{Konz2003Temperature, Jobez2014Cavity, Zhong2015Optically, Ma2021One-hour}. 
Since the hyperfine splitting of $\mathrm{^{153}Eu^{3+}}$ ions is larger than that of $\mathrm{^{151}Eu^{3+}}$ ions \cite{Timoney2013Single}, this isotope is chosen here to achieve an AFC memory with a wider bandwidth. The maximum bandwidth of the spin-wave storage \cite{Ma2021One-hour, Ma2021Elimination} can reach $29.1$ MHz for site-1 $\mathrm{^{153}Eu^{3+}}$ ions in the $\mathrm{Y_2SiO_5}$ crystal, which can be obtained based on the level structure of the $\mathrm{^{153}Eu^{3+}}$:$\mathrm{Y_2SiO_5}$ crystal (Supplementary Figure \ref{fig:Time sequence}\textbf{a}) and the analogous analytical method presented in Ref. \cite{SU2022ondemand}.

The detailed setup of quantum memory is shown in Supplementary Figure \ref{fig:quantum memory}.
To isolate mechanical vibrations and extend the optical coherence lifetime ($T_2$), the $\mathrm{^{153}Eu^{3+}}$:$\mathrm{Y_2SiO_5}$ crystal is cooled to 3.1 K with vibration-isolated sample holder installed in a closed-cycle cryostat (Montana Instruments). The sample holder uses five springs made from stainless steel with a wire diameter of 0.4 mm, which significantly reduces the high-frequency (above kHz) vibration caused by the cold head of the cryostat. $T_2$ is measured by the two-pulse photon echo \cite{Tittel2010Photon-echo}.
As shown in Supplementary Figure \ref{fig:T2-compare}, $T_2=1.27 \pm 0.01$ ms is measured with the vibration-isolated sample holder when the peak power of the input pulse is $1.71$ mW, which is nearly independent with time, while $T_2=0.49 \pm 0.04$ ms is measured at the moment of minimum vibration without this sample holder. 
The vibration of the cryostat is indirectly measured by an acceleration sensor placed near the cold head. 
Our vibration-isolated sample holder increases the $T_2$ of $\mathrm {^{153}Eu^{3+}}$ ions by more than 2.6 times, which is crucial for the preparation of high-resolution AFCs for long-lived and multiplexed quantum storage.

\begin{figure}[tbph]
	\includegraphics [width=0.6\textwidth]{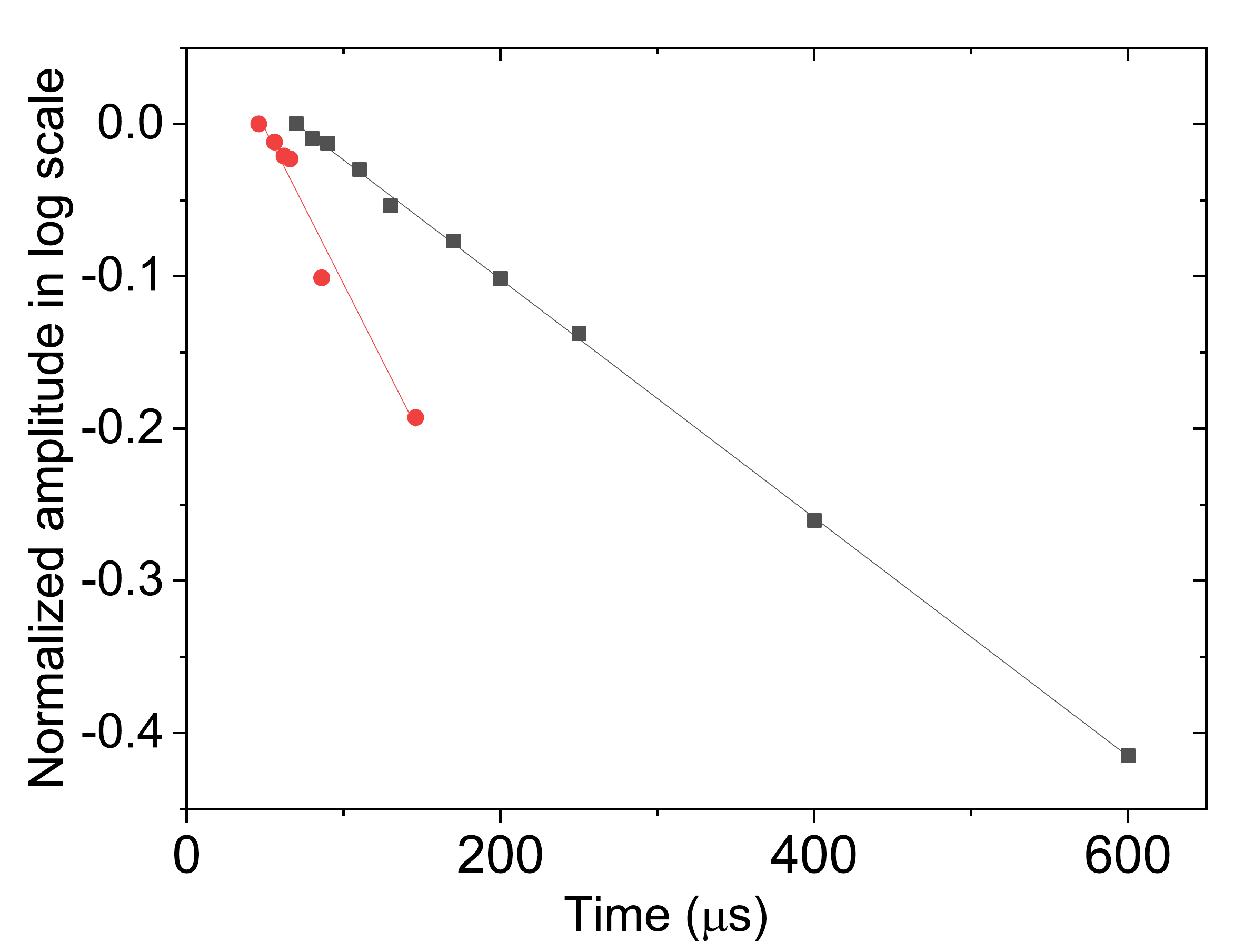}
	\caption{Measurements of the optical coherence lifetimes ($T_2$). The data measured with the vibration-isolated sample holder are indicated with black dots, while the data measured without the sample holder are presented with red dots. The horizontal axis represents the total storage time of the two-pulse photon echo. The solid lines are linear fittings, which give $T_2$ of $1.27 \pm 0.01$ ms with the vibration-isolated sample holder and $T_2 = 0.45 \pm 0.04$ ms without that. 
	}
	\label{fig:T2-compare}
\end{figure}

\begin{figure}[tbph]
	\includegraphics [width=1.0\textwidth]{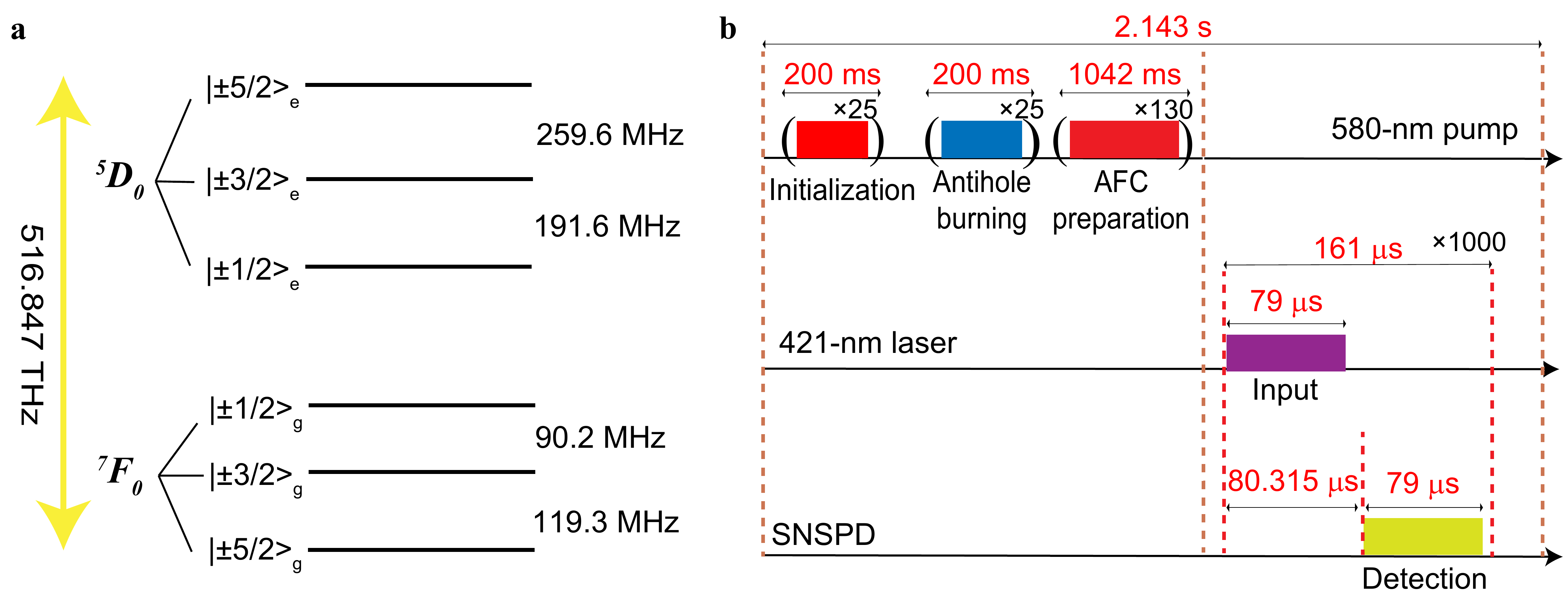}
	\caption{\textbf{a}, The level structure of the $\mathrm{{^7}F{_0}\rightarrow{^5}D{_0}}$ transition of the site-1 $\mathrm{^{153}Eu^{3+}}$:$\mathrm{Y_2SiO_5}$ crystal in the background of Earth’s magnetic field.	\textbf{b}, The time sequence. The preparation of the AFC memory takes 1.442 s. A single measurement cycle takes 161 $\mu$s and is repeated 1000 times. In each measurement cycle, the 421-nm laser is gated into pulses with a duration of 79 $\mu$s to generate the input photons in a pulsed fashion. The 1097 temporal modes are stored simultaneously within the input window of 79 $\mu$s. The SNSPD detects the retrieved photons after the storage time of 80.315 $\mu$s.}
	\label{fig:Time sequence}
\end{figure}

The input mode with a center frequency of $f_0 = 516.847$ THz is generated from the SPDC source, stored in the crystal for 80.315 $\mu$s, and finally released to the local gate $C_{21}$. We use two counter-propagating modes (pump-1 and pump-2) to prepare quantum memory. The pump pulses are generated by AOMs that are controlled with an arbitrary waveform generator.
Pump-1, modulated by a double-pass AOM, is applied for the initialization and the preparation of the AFC at the center frequency of $f_0$.
Pump-2, modulated by a single-pass AOM and a double-pass AOM, is employed to achieve antihole burning at the center frequency of $f_0$.
The time sequence for the preparation of quantum memory and the whole experiment is shown in Supplementary Figure \ref{fig:Time sequence}\textbf{b}. 

The first step of the pump sequence is initializing the absorption of the crystal. A chirping pulse at the center frequency of $f_0$ with a bandwidth of 68 MHz is repeated 25 times. After the initialization process, the population in the storage bandwidth is close to empty.

The next step is antihole burning. Based on the level structure of the $\mathrm{^{153}Eu^{3+}}$:$\mathrm{Y_2SiO_5}$ crystal (Supplementary Figure \ref{fig:Time sequence}\textbf{a}) and the analysis detailed in Ref. \cite{Zhu2022On-Demand}, an absorption band with an optical depth of $9.7$ at the center frequency of $f_0$ is obtained by applying a chirping pulse at the center frequency of $f_0-166$ MHz. This chirping pulse, which has a bandwidth of 158 MHz and a duration of 8 ms, is repeated 25 times. Using this pumping scheme, we can enhance the absorption depth by obtaining many back-burning holes from the different classes of ions in the absorption band. 

The third step is the preparation of the AFC. To achieve long-lived multiplexed storage, an AFC with an extremely large number ($N$) of comb teeth ($N>1000$) is required here. In the simple parallel preparation scheme \cite{Jobez2016Towards}, the preparation pulses are the direct superposition of the complex hyperbolic secant (CHS) pulses with periodic frequency detunings $n\Delta$, where $\Delta$ is the frequency period of the AFC and $n=1,2,3,...,N$. The interference between the CHS pulses leads to a low energy utilization of the AFC-preparation pulses.
As a result, the pulse energy of such simple parallel AFC-preparation pulses is proportional to $1/N$ \cite{Businger2022Non-classical}, which would require long preparation times to prepare AFCs with a large $N$, and the efficiency is severely limited. 
To solve this problem, Ref. \cite{Schroeder1970Synthesis, Oswald2021Burning, Businger2022Non-classical} proposed adding the Schroeder phase $\Phi_n=\pi\left[\frac{n^2}{2 N}\right]$ to every CHS pulses \cite{Schroeder1970Synthesis} to evenly separate the temporal centers of every CHS pulse \cite{Businger2022Non-classical}. Such an idea greatly improves the energy utilization efficiency of the AFC preparation pulse and leads to the preparation of high-quality AFCs \cite{Businger2022Non-classical}.
The CHS pulses $A_n(t)$ for square combs used here can be written as
\begin{equation}
	A_n(t) =\operatorname{sech}[\beta (t-\frac{T_{prep}}{2})] \sin \left[2 \pi\left(f_0+(n-\frac{N+1}{2}) \Delta\right) (t-\frac{T_{prep}}{2})+ 2\pi \frac{\Delta_f}{2 \beta} \ln (\cosh (\beta(t-\frac{T_{prep}}{2})\right], 
\end{equation}
where $t \in[0, T_{prep}]$, $n=1,2,3,...,N$, $T_{prep}$ is the pulse duration, parameter $\beta$ controls the waveform of the CHS pulses, $\gamma$ is the width of the combs and $\Delta_f=\Delta-\gamma$. 
By summing the CHS pulses with separated temporal centers $A'_n(t)$, the normalized AFC-preparation pulse $A(t)$ can be written as
\begin{equation}
	\begin{aligned}
	A(t) = &\frac{1}{\text{max}(A(t))} \sum_{n=1}^{N} \sin (\Phi_n) A'_n(t) \\
	= &\frac{1}{\text{max}(A(t))} \sum_{n=1}^{N} \sin (\pi\left[\frac{n^2}{2 N}\right])
	\left\{
	\begin{aligned}
		& A_n(t+\frac{n-1}{N}T_{prep})  & {0 \leq t < \frac{N-n+1}{N}T_{prep}} \\
		& A_n(t-\frac{N-n+1}{N}T_{prep})  & {\frac{N-n+1}{N}T_{prep} \leq t < T_{prep}} 
	\end{aligned}
	\right..
\end{aligned} 
\end{equation}
This waveform can only be applied to a single-pass AOM but not to a double-pass AOM. Here, we solve this problem by extending the function to the complex field that uses $\exp(i\theta(t))$ to replace the $\sin(\theta(t))$ term in $A(t)$. Considering the desired waveform 
\begin{equation}
		f(t) = a(t)e^{i\theta(t)} \quad \quad \text{with } a(t) \geq 0, \quad \theta(t) \in[-\pi, \pi],
\end{equation}
the corresponding waveform $f'(t)$ for the double-pass AOM can be represented as
\begin{equation}
\begin{aligned}
	f'(t) = & \sqrt{a(t)}e^{i\theta(t)/2} \quad \quad \text{with }\theta(t)/2 \in[-\pi/2, \pi/2],\\
	= & \sqrt{a(t)}e^{i\theta'(t)} \quad \quad \text{with }\theta'(t) \in[-\pi, \pi]. 
\end{aligned} 
\end{equation}  
Since $\theta(t)/2$ is a discontinuous function, it is necessary to perform sign inversion on part of $\theta(t)/2$ to make the function continuous, which is denoted as $\theta'(t)$.

\begin{figure}[tbph]
	\includegraphics [width=0.9\textwidth]{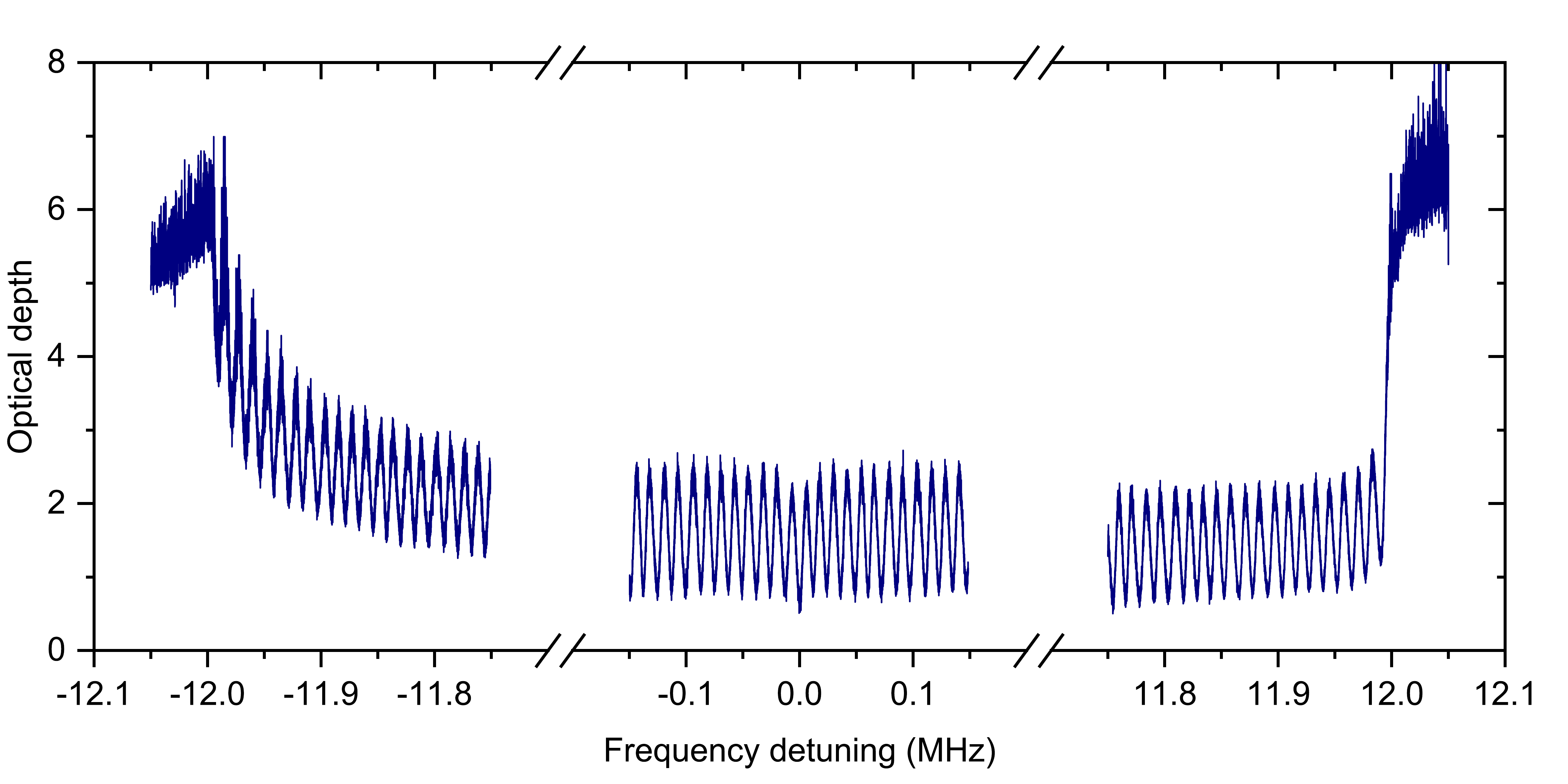}
	\vspace{-0.3cm}
	\caption{The measured structure of the prepared atomic frequency comb (AFC), with a storage time of 80.315 $\mu$s and a total bandwidth of 24 MHz. The AFC comprises a total of 1927 comb teeth; however, only three subsets are measured and depicted to ensure the structure is clearly visible.}
	\label{fig:AFC}
	\vspace{0.2cm}
\end{figure}

The optimized AFC-preparation pulse in our experiment is obtained with parameters $T_{prep}=8.018$ ms, $\beta=17.627/T_{prep}$, and $\Delta_f=\Delta/11$. The pulse is repeated 130 times, and the peak power of the AFC-preparation pulse is 142 $\mu$W before the cryostat. Finally, the 24-MHz AFC structure (Supplementary Figure \ref{fig:AFC}) is successfully prepared in the $\mathrm{^{153}Eu^{3+}}$ ion ensemble. 
Limited by the accuracy of comb preparation, the non-zero absorption between the AFC teeth results in a decrease in efficiency.

As shown in Fig. 1\textbf{c} in the main text, the storage efficiency for 580-nm entangled photons is measured by coincidence counts triggered by paired 1537-nm photons. To reduce the time interval between the entangled photon pairs for more accurate measurement results, the data after 80.315-$\mu$s AFC storage (the blue line in Fig. 1\textbf{c}) is measured by delaying the 1537-nm entangled photons with the 7.9-km field-deployed fiber loop. The emergence of the additional small peak in the blue traces, as shown in Fig. 1\textbf{c} in the main text, can be attributed to the bandwidth of the input photons exceeding that of the AFC memory. Since the storage bandwidth of the AFC memory is square-enveloped, its Fourier transform in the temporal domain is a sinc function. This results in additional small peaks after the main peak in the temporal domain.

The bandwidth of 580-nm entangled photons is 178 MHz, which is still greater than the 24-MHz bandwidth of the memory. Therefore, the AFC memory essentially acts as a spectral filter for the quantum light source. We measure the input and the transmission of the quantum light, by preparing the absorption to a 24-MHz transparent window and a 24-MHz AFC, respectively (red and black lines in Fig. 1\textbf{c}). The random coincidences have been subtracted to accurately determine the storage efficiency for the quantum light at the bandwidth of 24 MHz. The measured AFC efficiency decays exponentially with the storage time with an effective AFC lifetime $T_{2}^{AFC} = 148 \pm 6$ $\mu$s, as shown in Supplementary Figure \ref{fig:mode_number_AFC_decay}\textbf{a}.  The storage time ($1/e$ intensity decay time) of the 24-MHz memory is 37 $\mu$s. It is worth noting that a decrease in efficiency was observed with the expansion of the AFC bandwidth, which may be ascribed to the effect of spectral diffusion. 
For the bandwidth-matched 580-nm photons, the efficiency is $3.2 \pm 0.1 \%$ for an 80.315-$\mu$s storage.

\begin{figure}[tbph]
	\includegraphics [width=1\textwidth]{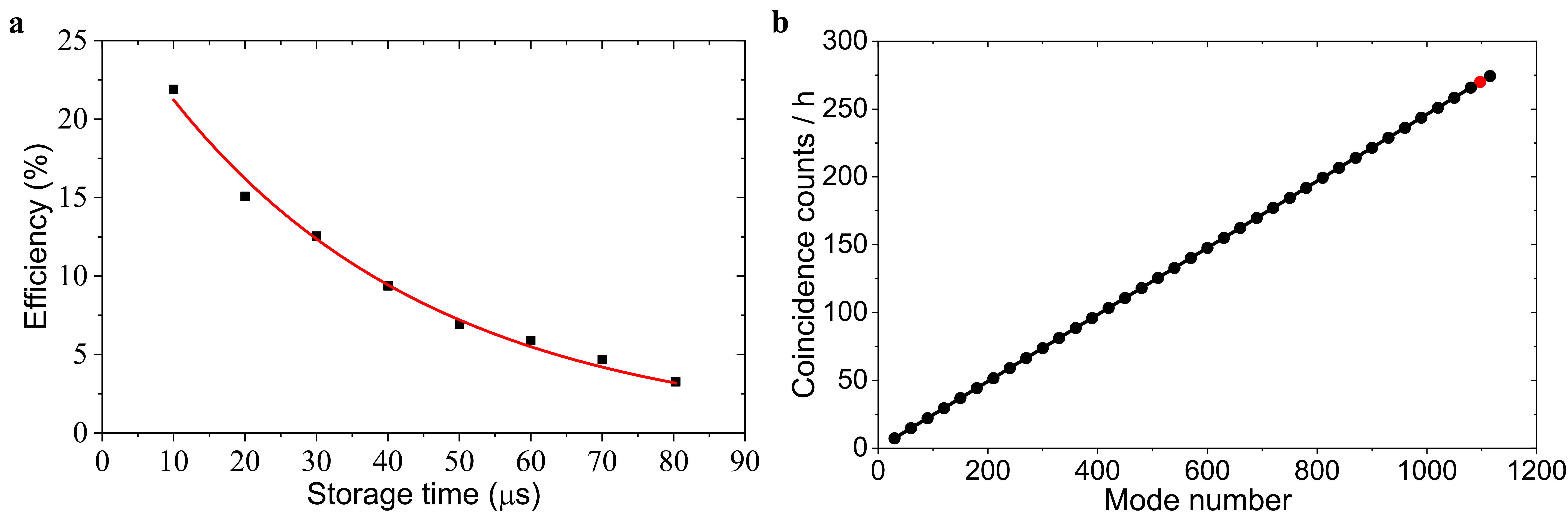}
	\caption{ \textbf{a}, The storage efficiency as a function of the storage time with a bandwidth of 24 MHz. The red line is fitted by the formula $\eta_0 \exp(-4/T_{2}^{AFC})$ \cite{Jobez2016Towards}, resulting in $T_{2}^{AFC} = 148 \pm 6$ $\mu$s and $\eta_0 = 27.8 \pm 0.9 \%$. \textbf{b}, The average coincidence counts per hour as a function of the number of temporal modes. The red dot indicates 1097 temporal modes as used in the actual experiments. The black dots are estimations from the experimental data.
 }
	\label{fig:mode_number_AFC_decay}
\end{figure}

The storage time demonstrated here is already close to the limit of excited-state storage, and longer storage times can be expected with spin-wave storage \cite{Ma2021One-hour,Ma2021Elimination}. 
We would need to choose an appropriate lambda system and select out one class of ions for implementing spin-wave storage. For the $\mathrm{^{153}Eu^{3+}}$:$\mathrm{Y_2SiO_5}$ crystals in the background of the Earth’s magnetic field, if we use a similar lambda system and apply a similar methodology as in Ref. \cite{SU2022ondemand} to design the pumping sequence, the spin-wave storage bandwidth can reach 29.1 MHz. This lambda system contains the $|\pm1/2\rangle_g \rightarrow |\pm5/2\rangle_e$ transition for the absorption of signal photons, the $|\pm5/2\rangle_g \rightarrow |\pm5/2\rangle_e$ transition for the control field, and the auxiliary $|\pm3/2\rangle_g \rightarrow |\pm3/2\rangle_e$ transition for class-cleaning and spin-polarization.
Due to the extended spin coherence time of $\mathrm{Eu^{3+}}$ in $\mathrm{Y_2SiO_5}$ crystals, the storage time for spin-wave quantum memory has been demonstrated to reach several tens of milliseconds \cite{Ortu2022Storage}, with an upper limit extending up to a few hours \cite{Zhong2015Optically,Ma2021One-hour}, which could lead to a quantum storage performance far beyond the capabilities of fiber delay loops. Moreover, spin-wave quantum memory provides on-demand retrieval \cite{Jobez2015Coherent} and enables real-time manipulation of photonic qubits \cite{Yang2018Multiplexed}, which are crucial for quantum repeaters and large-scale quantum networks \cite{Lei2023Quantum}.

We can define a single mode duration of 72 ns that completely covers the retrieved echo, there are 79 $\mu$s / 72 ns = 1097 temporal modes simultaneously stored in the memory crystal \cite{Lago-Rivera2021qp,liu2021heralded, Businger2022Non-classical}. As shown in Supplementary Figure \ref{fig:mode_number_AFC_decay}\textbf{b}, temporal multimode storage enables a linear enhancement of the rate for quantum gate teleportation.
In our implementation, the multimode capacity surpasses the performance of other ensemble-based quantum memories, such as cold atoms \cite{Parniak2017Wavevector,Pu2017Experimental} and hot vapor \cite{Chrapkiewicz2017High}. However, the storage efficiency is currently lower than that of cold atoms and hot vapor \cite{Wang2019Efficient,Ma2022High}. This is primarily caused by the weak absorption of the $\mathrm{^{153}Eu^{3+}}$:$\mathrm{Y_2SiO_5}$ crystals, and the 54\% theoretical limit posed by the AFC memory for retrieval in the forward direction \cite{Afzelius2009Multimode}. To address this limitation, cavity-enhanced AFC memory could be utilized to improve storage efficiency \cite{Afzelius2010Impedance,duranti2023efficient}.

\subsection{6. Optical losses}

Due to the low duty cycle of memory operations (3.7\%), the coincidence count rate of the entire experiment is approximately 0.042 Hz. After quantum storage, the retrieved echo has a signal-to-noise ratio of 12.6:1. The data collection times for the Deutsch-Jozsa algorithm and quantum phase estimation algorithm in our experiment were 3.8 hours and 17.1 hours, respectively.

The collection efficiency of the unfiltered SPDC source is approximately $42\%$. The heralding efficiency of the narrowband SPDC sources is $4.9\%$. The main optical losses are listed here. The peak transmittance of the cascaded etalons is approximately 63\%, both for 580 nm and 1537 nm. The detection efficiency is $80\%$ and $91\%$ for photons at 580 nm and 1537 nm, respectively. The 7.9 km field-deployed optical fiber introduces a loss of 2.2 dB, which includes a transmission loss of 1.6 dB and a coupling loss of 0.6 dB. The whole fiber channel can maintain a polarization extinction ratio of greater than 100:1 for 24 hours (see Supplementary Figure \ref{fig:polarization}).
 
The source generates energy-time entanglement, the interferometers select two time bins from the general superposition and each unbalanced interferometer introduces postselection and coupling loss. The coupling efficiency is $71\%$ for 580 nm and $62\%$ for 1537 nm.

The internal storage efficiency for 24-MHz quantum light is $3.2\%$ and the total transmission of the optical setup for the quantum memory is approximately $65\%$. In addition, there is a bandwidth match between the quantum light source and the quantum memory, and approximately $8\%$ of the generated photons are in the bandwidth of the memory.

The success probabilities or efficiencies of each step and the losses of each component are also summarized in Supplementary Table \ref{tabloss}.

\begin{table}
    \centering
    \renewcommand\arraystretch{1.22}
    \begin{tabular}{|c|c|}
\hline \textbf{Steps and components} & \textbf{Success probabilities or losses} \\
\hline
Success probability of the gate teleportation scheme & $50\%$\\
\hline
The postselection of unbalanced interferometer (for 580 nm or 1537 nm) & $50\%$\\
\hline 
Collection efficiency of the unfiltered SPDC source &  $42\%$ \\
\hline
Heralding efficiency of the narrowband SPDC sources & $4.9\%$\\
\hline
Peak transmittance of the cascaded etalons (for 580 nm or 1537 nm) & $63\%$\\
\hline
Loss of the 7.9 km field-deployed optical fiber & 2.2 dB\\
\hline
Coupling efficiency of the unbalanced interferometer (for 580 nm) & $71\%$\\
\hline
Coupling efficiency of the unbalanced interferometer (for 1537 nm) & $62\%$\\
\hline
Internal storage efficiency quantum memory for 24-MHz quantum light & $3.2\%$\\
\hline
Total transmission of the optical setup for quantum memory & $65\%$\\
\hline
Bandwidth matching efficiency between photons and the quantum memory & $8\%$\\
\hline
Detection efficiency (for 580 nm) & $80\%$\\
\hline
Detection efficiency (for 1537 nm) & $91\%$\\
\hline
\end{tabular}
    \caption{The success probabilities or efficiencies of each step and the losses of each component.}
    \label{tabloss}
\end{table}

\begin{figure}[tbph]
\includegraphics [width=0.6\textwidth]{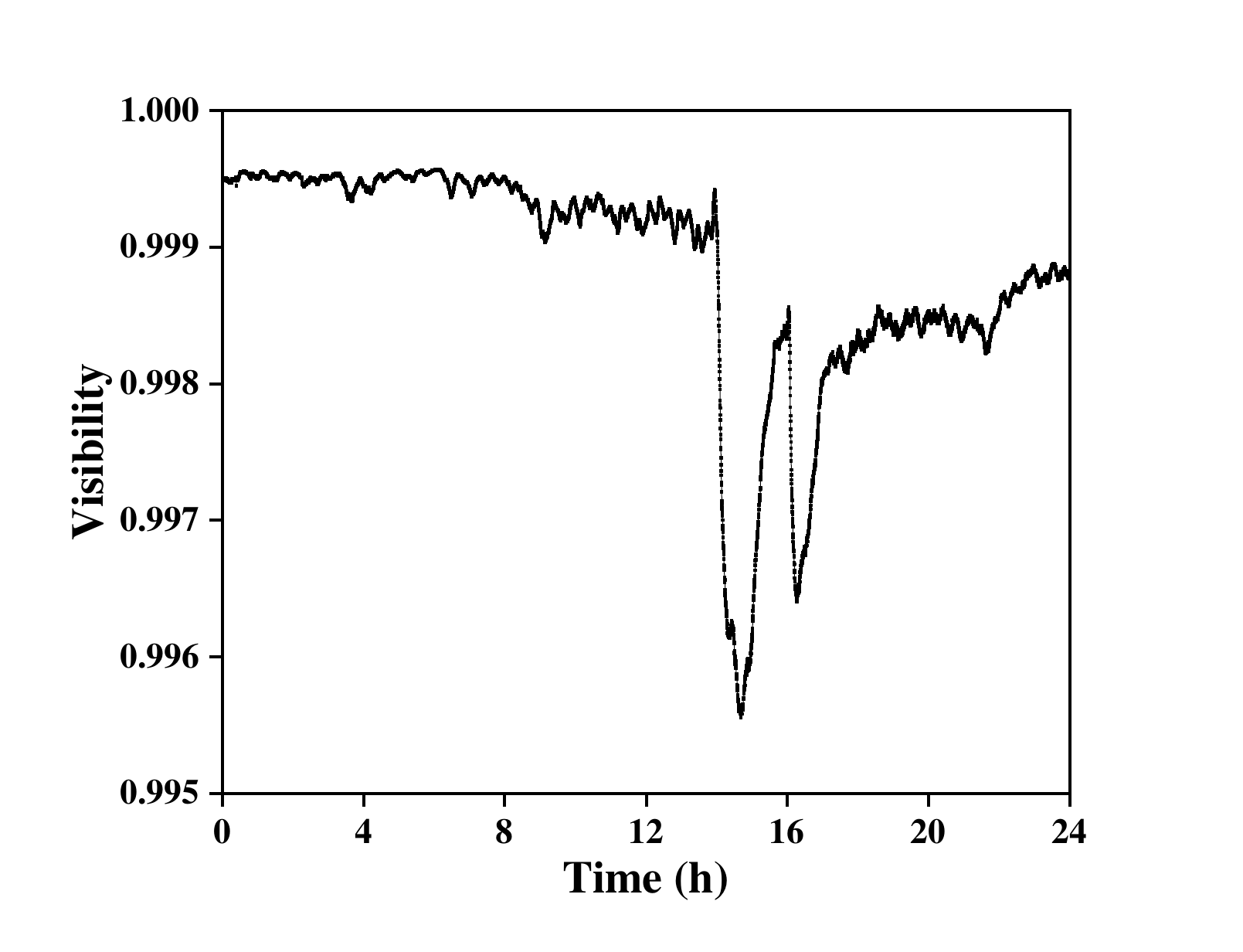}
\caption{Monitoring of polarization stability of field-deployed optical fiber. The vertical axis represents the visibility of the horizontal polarization state. Here $\text{visibility} = (n(H)-n(V))/(n(H)+n(V))$ with $n(H,V)$ is the intensity of measured light with corresponding polarization states. The visibility of polarization states after transmission over the entire fiber channel is maintained above 0.995 over 24 h.}
\label{fig:polarization}
\end{figure}

\subsection{7. Synchronization of the entanglement distribution system }

\begin{figure}[tbph]
\includegraphics [width=0.8\textwidth]{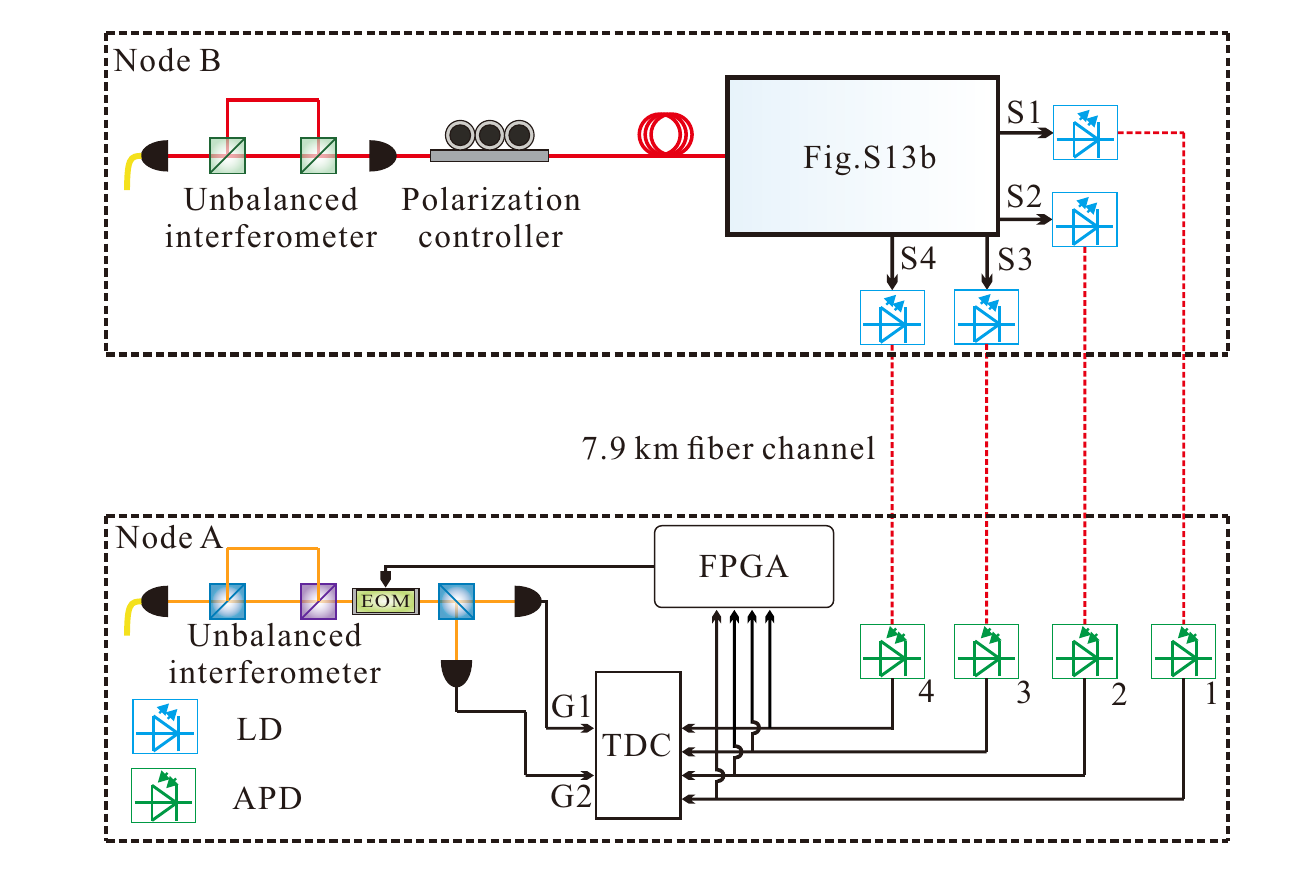}
\caption{Synchronization system between node A and node B. The classical channel and quantum channel use different fibers from the same fiber-optic cable, with a length of 7.9 km. In node A, 580-nm photons are detected using SNSPD to obtain pulse signals G1 and G2. Similarly, in node B, 1537-nm photons are also detected using SNSPD to obtain four pulse signals, S1, S2, S3, and S4. To achieve synchronization between node A and node B, the electrical signal pulse from node B is converted into optical signal pulses through a laser diode (LD) and transmitted to node A through classical channels. At node A, the avalanche photodiode (APD) converts the optical pulse back into an electrical signal pulse. The signals S1, S2, S3, S4, G1, and G2 are sent into a time-to-digital converter (TDC) to achieve time-digital conversion, allowing the system to obtain the coincidence counts between node A and node B. The signals S1, S2, S3, and S4 are further processed by a field programmable gate array (FPGA), and the resulting logic signals are applied to an EOM to implement local operations.}
\label{fig: synchronous}
\end{figure}

We achieve synchronization between node A and node B through electro-optic and photoelectric conversion (see Supplementary Figure \ref{fig:synchronous}), with a synchronization accuracy of 220 ps. At node B, the four signals (S1, S2, S3 and S4) from the SNSPD are converted into optical pulses at wavelengths of 1550.92 nm, 1550.12 nm, 1549.32 nm, and 1548.52 nm, respectively.
These optical pulses are then transmitted through a 7.9-km optical fiber to node A, where they are converted back into electrical signals using avalanche photodiodes (APDs). The time-to-digital converter (TDC) is utilized to convert the time tag signals from both node A (G1, G2) and node B (S1, S2, S3, S4) into digital signals, enabling coincidence readout. To realize LOCC, a field programmable gate array (FPGA) is employed to determine the appropriate operation based on the signal received from node B. The electro-optical modulator (EOM) in node A performs $\sigma_x$ operation when receiving signals from S1 and S2, and $I$ operation when receiving signals from S3 and S4.

\subsection{8. Unbalanced interferometer locking}

Here, we will provide a detailed description of the phase locking methods used for both the 1537-nm and 580-nm unbalanced interferometers (see Supplementary Figure \ref{fig:lock_phase}). We use a single-photon-level reference light to lock the phase of the unbalanced interferometer. The reference light counter-propagates with the signal light to avoid introducing too much noise. Phase compensation is realized by a piezoelectric transducer (PZT), which drives a mirror to move in free space. All fibers in the interferometer are equipped with FC-APC connectors to minimize reflections. We must ensure $\tau_{\text {delay}} \gg \tau_{\text {photon}}$, where $\tau_{\text {delay}}$ corresponds to the time delay caused by the difference between the long and short arms, and $ \tau_{\text {photon}}$ corresponds to the coherence time of single photons, which is related to the photon bandwidth (155 MHz for 1537-nm photons and 24 MHz for 580-nm photons). Therefore, we use 30-m long optical fibers to build unbalanced interferometers to meet such conditions. For effective phase locking, we carefully select the reference light to have the same wavelength as the signal light.

\begin{figure}[tbph]
\includegraphics [width=1\textwidth]{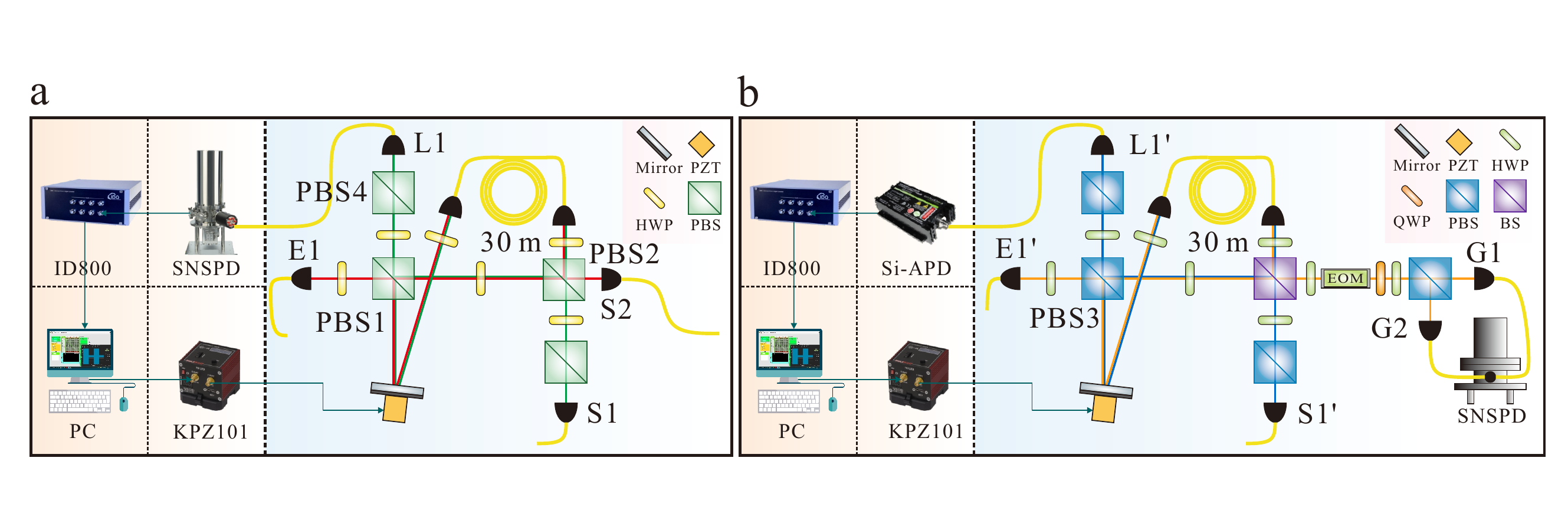}
\caption{Phase locking of the unbalanced interferometers. \textbf{a}, Phase locking at 1537 nm. The reference light and the signal light travel in opposite directions. The 1537-nm reference light (green lines) with a polarization of $(|H\rangle+|V\rangle)/\sqrt{2}$ is emitted from S1 and transformed into path coding by PBS2. It then enters the long (30-m optical fiber) and short (free space between PBS1 and PBS2) arms with $H$ and $V$ polarization components, respectively. The unbalanced interferometer combines the light from the long and short arms, resulting in $(|H\rangle+e^{i\phi}|V\rangle)/\sqrt{2}$ on PBS1. The reference light is then projected into $(|H\rangle-|V\rangle)/\sqrt{2}$ basis on PBS3 and analyzed by a SNSPD and a time-to-digital converter (IDQ, ID800). The feedback signal is amplified by a high-voltage controller (Thorlabs, KPZ101), and applied to the piezoelectric transducer (PZT) to stabilize the phase. \textbf{b}, Phase locking at 580 nm. This module is only slightly modified compared to that shown in \textbf{a}. The PBS2 in \textbf{a} is replaced by a beamsplitter (BS) to implement the path measurement at 580 nm. The phase-locking strategy is the same, and here, Si-based APD is employed to detect the reference light at 580 nm.}
\label{fig:lock_phase}
\end{figure}

Without phase locking, the phase of the unbalanced interferometer varies by 2$\pi$ at a rate of approximately 0.1 Hz. We set the PZT locking frequency to 10 Hz to lock the interferometer and the phase locking can be stable for over 24 hours. 
The background noise finally introduced by the reference light is less than 100 Hz and 6 Hz for 580 nm and 1537 nm, respectively, which has negligible effects on the signal-to-noise ratio in our experiments.

\subsection{9. Encoding converter and local CNOT operation, C-\textit{U} operation}

Energy conservation guarantees that two photons in a pair are generated at the same time, and the generation time of the photon pair is uncertain within the coherence time of the pump laser, resulting in time-energy entanglement \cite{franson1989bell,clausen2011quantum}. The time-energy entanglement is ensured as long as $\tau_{\text {pump }} \gg \tau_{\text {pair }}$, where $\tau_{\text {pump}}\left(\tau_{\text {pair}}\right)$ corresponds to the coherence time of the pump laser light (photon pair). As shown in Supplementary Figure \ref{fig:Encoding converter}, we divide photons into long paths and short paths with equal probability through HWPs (HWP1 and HWP3 ) set to 22.5 degrees and PBSs (PBS1 and PBS3). We generate long ($|L\rangle$) and short ($|S\rangle$) paths using a 30-m unbalanced interferometer. At this point, the entangled state of the two photons is

\begin{align}
    |\psi\rangle=1/\sqrt{2}(|S\rangle_2|S\rangle_3+|L\rangle_2|L\rangle_3).
\end{align}

We also use PBS2 to convert $|S\rangle$ into $|H\rangle$ and $|L\rangle$ into $|V\rangle$ at 1550 nm. The entangled state of the two photons thus becomes

\begin{align}
    |\psi\rangle=1/\sqrt{2}(|S\rangle_2|H\rangle_3+|L\rangle_2|V\rangle_3).
\end{align}

Afterwards, we distribute polarized photons to node B through an external field fiber.
At nodes A and B, we encode qubits 1, 2 and 3, 4 using photons with different degrees of freedom, and implement $C_{21}$ and $C_{43}$ operations, respectively.

\begin{figure}[tbph]
\includegraphics [width=0.6\textwidth]{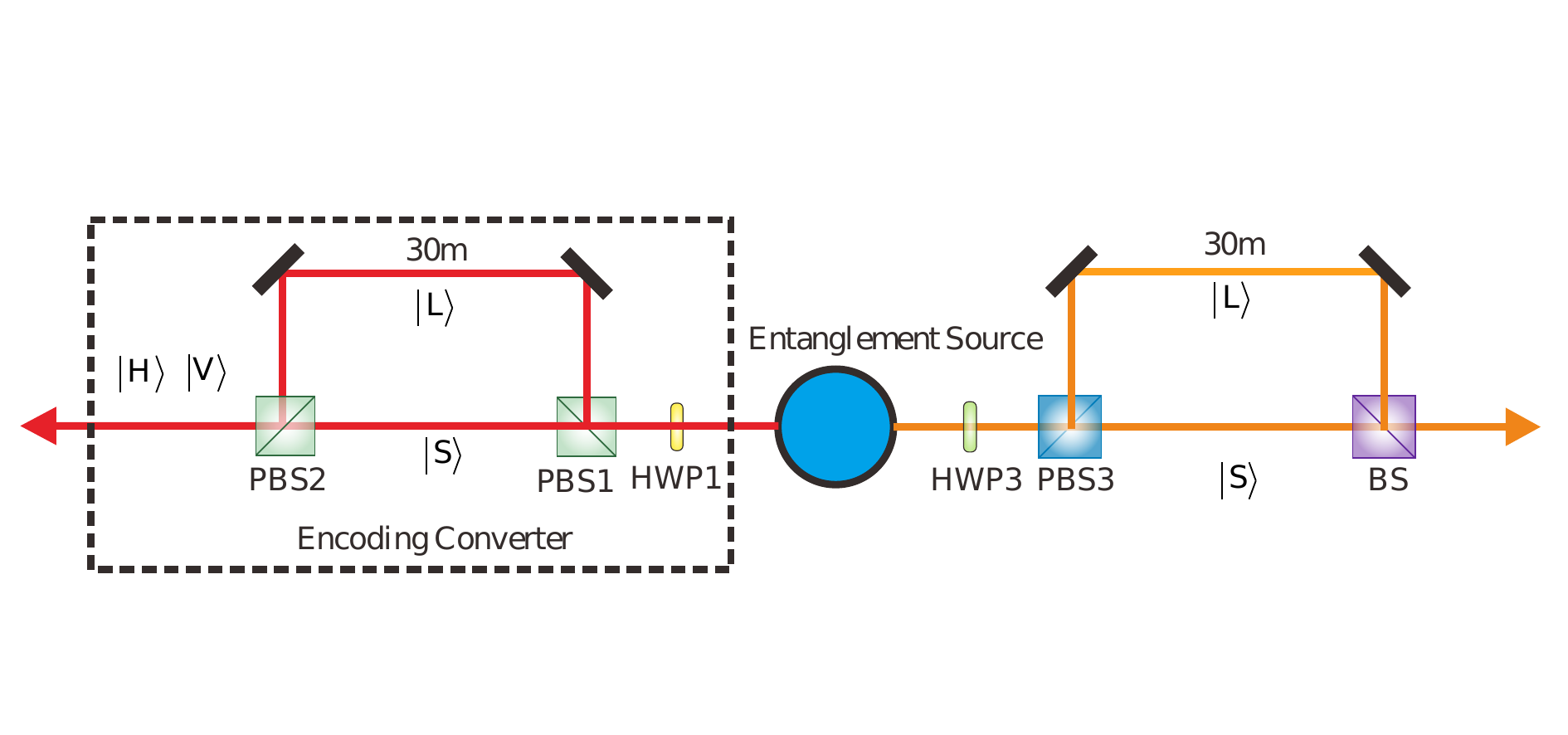}
\caption{Encoding converter. There are two unbalanced interferometers at 1550 nm and 580 nm, respectively, forming the postselection of time-energy entanglement. PBS3 is used to achieve the conversion of time-energy DOF and polarization DOF. }
\label{fig:Encoding converter}
\end{figure}

In Supplementary Figure \ref{fig:C-NOT_of_580nm_and_1537nm}, at node B, we convert the polarization DOF into path DOF through BD1. At the same time, we define $|S\rangle$ and $|L\rangle$ as path DOFs $|0\rangle$ and $|1\rangle$ at node A, and the entangled state becomes path-entangled state:

\begin{align}
    |\psi\rangle=1/\sqrt{2}(|0\rangle_2|0\rangle_3+|1\rangle_2|1\rangle_3).
\end{align}

In Supplementary Figure \ref{fig:C-NOT_of_580nm_and_1537nm}, wave plates are employed to encode qubit 1 and qubit 4 in the polarization DOF of two photons. The quantum state of the whole system can be written as
\begin{align}
    |\psi\rangle_{1234}=(\alpha_1|H\rangle+\beta_1|V\rangle)_1[(|0\rangle|0\rangle+|1\rangle|1\rangle)_{23}/\sqrt{2}] (\alpha_4|H\rangle+\beta_4|V\rangle)_4, \label{qubit1234}
\end{align}
where the subscripts represent different qubits. Qubits 1 and 2 are carried by photons at 580 nm, while qubits 3 and 4 by photons at 1537 nm.
 
In Supplementary Figure \ref{fig:C-NOT_of_580nm_and_1537nm}\textbf{a}, we directly encode the path and polarization qubits 1 and 2 in the unbalanced interferometer. In node A, we implement the CNOT operation of the path qubit on the polarization qubit by setting an HWP of $45^{\circ}$ in path 1 (if we want to achieve C-\textit{U} operation of the path to polarization DOF, we only need to use the combination of HWPs and QWPs to achieve \textit{U}-operation of polarization DOF in path 1). Afterward, we use polarization-independent beamsplitters to achieve measurements of the path DOF on $(|0\rangle+|1\rangle)/\sqrt{2}$ and $(|0\rangle-|1\rangle)/\sqrt{2}$ bases.

\begin{figure}[tbph]
\includegraphics [width=1\textwidth]{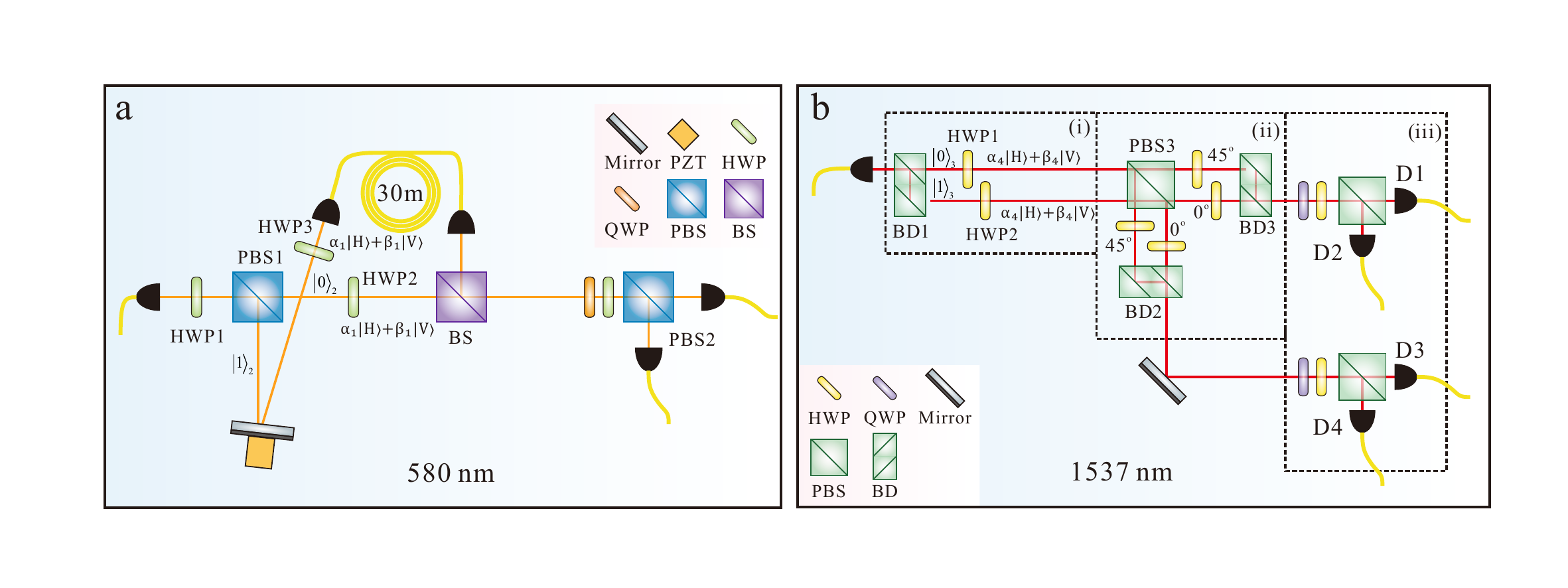}
\caption{Implementations of local CNOT gates.
\textbf{a}, Path-polarization CNOT gate at node A. PBS1 converts qubit 2 from time-energy encoding to path encoding ($|0\rangle_{2}$, $|1\rangle_{2}$). The polarization quantum state $\alpha_1|H\rangle+\beta_1|V\rangle$ encoding on each path is realized through HWP2 and HWP3. Then, the CNOT operation of the path DOF on the polarization DOF is achieved by rotating HWP3 by an additional $45^{\circ}$ on path 1. BS is used to achieve $(|0\rangle+|1\rangle)/\sqrt{2}$ and $(|0\rangle-|1\rangle)/\sqrt{2}$ measurements of the path DOF. In our scheme, we only retained the measurement results of $(|0\rangle+|1\rangle)/\sqrt{2}$ and discarded the measurement results of $(|0\rangle-|1\rangle)/\sqrt{2}$. Finally, the polarization state is analyzed using QWP, HWP, and PBS. \textbf{b}, Path-polarization CNOT gate at node B. (i) Preparation of path and polarization states. A beam displacer, BD1 converts the polarization state of qubit 3 into a path state ($|0\rangle_{3}$, $|1\rangle_{3}$). HWP1 and HWP2 prepare the polarization state ($\alpha_4|H\rangle+\beta_4|V\rangle$) on paths 0 and 1, respectively. (ii) Polarization-path CNOT gate and path-state measurement. HWP, PBS3, BD2, and BD3 together achieve the CNOT operation. When qubit 4 is in the $|H\rangle$ state, the path state remains unchanged, and when it is in the $|V\rangle$ state, the path state flips.
(iii) Standard polarization-state analysis. }
\label{fig:C-NOT_of_580nm_and_1537nm}
\end{figure} 

In Supplementary Figure \ref{fig:C-NOT_of_580nm_and_1537nm}\textbf{b}, we use a stable interferometer based on beam displacers (BD) to achieve path and polarization qubit encodings and operations. This interferometer is robust to noise and is suitable for out-field laboratories with poor conditions. In node B, we implement the CNOT operation of the polarization qubit on the path qubit.  
If the polarization is $H$, the path state remains unchanged, and when the polarization state is $V$, the path state flips.

\subsection{10. The implementation of the Deutsch-Jozsa and phase estimation algorithm}

Here we will provide a detailed introduction to the experimental implementation of these two algorithms. The Deutsch-Jozsa and phase estimation algorithms rely on the constructed nonlocal CNOT and C-\textit{U} gates, respectively. In the Deutsch-Jozsa algorithm, as shown in Fig. 3\textbf{a} of the main text, we implement four operations between A$_1$ and B$_4$: identity (ID) and NOT operations represent the constant function; CNOT and zero-CNOT (ZCNOT) operations represent the balanced function. For identity (ID) and NOT operations, there is no need for the A$_1$ qubit to perform nonlocal two-qubit operations on the B$_4$ qubit. ID operation does not require any operations on A$_1$ and B$_4$, and NOT operation only requires a bit flip operation on B$_4$ (HWP@$45^{\circ}$). Entangled resources are not needed for the constant functions, so we directly distribute two photons that are in the direct product of polarization DOF to A$_1$ and B$_4$, thus achieving the two functions. For CNOT and ZCNOT operations, we can use the nonlocal CNOT operations introduced earlier. ZCNOT operation requires performing bit flipping on B$_4$ based on the CNOT gate. In the Deutsch-Jozsa algorithm, the Fourier transform ($F$) can be achieved by setting the HWP at $22.5^{\circ}$ for polarization qubit. It is necessary to perform Fourier transform and inverse Fourier transform ($F^{-1}$) on the inputs and outputs of A$_1$ and B$_4$.

The quantum phase estimation algorithm is used to estimate the phase of an operator acting on an eigenstate and is frequently used as a subroutine in other quantum algorithms, such as factorization. In this algorithm, the quantum state register consists of a unitary operator $U$ with an eigenstate $|\psi\rangle$ ($U|\psi\rangle=\mathrm{e}^{\mathrm{i}2\pi \varphi}|\psi\rangle$), and information about the unitary operator $U$ is encoded on the measurement register through multiple controlled-$U^{2^{k}}$ operations with $k$ an integer. In the phase estimation algorithm, to calculate the probability of successfully determining each bit correctly, we initially assume that the phase, denoted by $\varphi$, can be expressed as a binary number with no more than $m$ bits: $\varphi=(0.\varphi_1\varphi_2...\varphi_m000...)$. During the first iteration ($k=m$), a controlled-$U^{2^m}$ gate is applied, targeting the $m$th bit of the expansion. The probability of measuring ``0" is $P_0 = \cos^2\left[\pi\left(0.\varphi_{m-1}00\ldots\right)\right]$, which equals unity when $\varphi_m = 0$ and zero when $\varphi_m = 1$. Consequently, the first bit $\varphi_m$ is extracted deterministically. In the subsequent iteration ($k=m-1$), the measurement focuses on the $(m-1)$th bit. The phase of the first qubit before the $Z$ rotation equals $2\pi\left(0.\varphi_{m-1}\varphi_m00\ldots\right)$. If we denote the first $m$ bits of the binary expansion of $\varphi$ as $\widetilde{\varphi} = 0.\varphi_1\varphi_2\ldots\varphi_m$, there generally exists a remainder $\delta$, where $\delta<1$, defined by $\varphi = \widetilde{\varphi} + \delta 2^{-m}$. Under these conditions, the probability of measuring $\varphi_m$ is $\cos^2(\pi \delta / 2)$. If $\varphi_{m}$ is measured correctly, the probability of accurately measuring $\varphi_{m-1}$ in the second iteration becomes $\cos^2(\pi \delta / 4)$, and so forth.

The experimental setup is similar to that of the Deutsch-Jozsa algorithm, as illustrated in Fig. 3\textbf{f} in the main text. According to Eq. \ref{CU}, in order to achieve a $50\%$ success probability of C-\textit{U} gate using LOCC, we use A$_1$ to control B$_4$ instead of B$_4$ to control A$_1$. By employing the C-\textit{U} gate and Fourier transform mentioned above, we can achieve a single-cycle phase estimation algorithm. We use a liquid crystal phaser to provide feedback on the phase information measured in each cycle ($Z_{\varphi}$). After 3 cycles, we can implement a 3-bit binary phase estimation algorithm.

\subsection{11. Detailed experimental results}

Here, we provide more detailed data about the experimental results presented in the main text.

The truth table of the remote CNOT gate is detailed in the Supplementary Table \ref{table1}.

The real parts of the key matrix elements for four Bell states generated by the remote CNOT gates are provided in Eq. \ref{eq1}-\ref{eq4}.
\begin{equation}
|\Phi^+\rangle=\left(\begin{array}{rrrr}
0.429 & 0 & 0 & 0.376 \\
0 & 0.064 & 0 & 0 \\
0 & 0 & 0.064 & 0 \\
0.376 & 0 & 0 & 0.443
\end{array}\right),
\label{eq1}
\end{equation}

\begin{equation}
|\Phi^-\rangle=\left(\begin{array}{rrrr}
0.472 & 0 & 0 & -0.406 \\
0 & 0.074 & 0 & 0 \\
0 & 0 & 0.037 & 0 \\
-0.406 & 0 & 0 & 0.417
\end{array}\right),
\label{eq2}
\end{equation}

\begin{equation}
|\Psi^+\rangle=\left(\begin{array}{rrrr}
0.056 & 0 & 0 & 0 \\
0 & 0.400 & 0.360 & 0 \\
0 & 0.361 & 0.504 & 0 \\
0 & 0 & 0 & 0.040
\end{array}\right),
\label{eq3}
\end{equation}

\begin{equation}
|\Psi^-\rangle=\left(\begin{array}{rrrr}
0.077 & 0 & 0 & 0 \\
0 & 0.505 & -0.357 & 0 \\
0 & -0.357 & 0.385 & 0 \\
0 & 0 & 0 & 0.034
\end{array}\right).
\label{eq4}
\end{equation}

The detailed data of the Deutsch-Jozsa algorithm are shown in the Supplementary Table \ref{table2}.

Detailed data for the phase estimation algorithm are shown in the Supplementary Table \ref{table3}.

\begin{table}[H]
\centering 
\begin{tabular}{ccccc}
\hline & $HH$ & $HV$ & $VH$ & $VV$ \\
\hline $HH$ & $0.894\pm 0.021$ & $0.106\pm0.020$ & 0 & 0 \\
\hline $HV$ & $0.106\pm 0.021$ & $0.894\pm0.020$ & 0 & 0 \\
\hline $VH$ & 0 & 0 & $0.129\pm0.023$ & $0.889\pm0.021$ \\
\hline $VV$ & 0 & 0 & $0.871\pm0.023$ & $0.111\pm0.021$ \\
\hline
\end{tabular}
\caption{Experimental data of truth table the CNOT gate. The horizontal axis represents the input, and the vertical axis represents the output. 0 represents a photon detection probability of 0 for this setting of measurements in the experiment. Compared to the theoretical truth table, the CNOT gate has a high fidelity of $0.887\pm0.021$.}
\label{table1}
\end{table}

\begin{table}[H]
\begin{equation}
\begin{array}{|c|cccc|}
\hline & \text { ID } & \text { NOT } & \text { CNOT } & \text { ZCNOT } \\
\hline H & 0.987 & 0.993 & 0.061 & 0.086 \\
\hline V & 0.013 & 0.007 & 0.939 & 0.914 \\
\hline \text {Errors} & 0.009 & 0.007 & 0.009 & 0.009 \\
\hline
\end{array}
\nonumber
\end{equation}
\caption{Experimental data of the Deutsch-Jozsa algorithm. The table gives the probabilities for measuring the polarization state of $H$ and $V$. The horizontal axis represents the four operations, ID, NOT, CNOT, and ZCNOT. The probabilities that the results are either $H$ or $V$ determine whether the function is balanced or constant. It can be seen that for each type of function, we can make a correct judgment with a probability of at least 0.91.}
\label{table2}
\end{table}

\begin{table}[H]
\begin{equation}
\begin{array}{|c|ccc|ccc|ccc|ccc|}
\hline & 0 & 0 & 0 & 0 & 1 & 0 & 1 & 0 & 1 & 1 & 1 & 0 \\
\hline 0 & 0.876 & 0.883 & 0.913 & 0.861 & 0.097 & 0.812 & 0.182 & 0.828 & 0.153 & 0.162 & 0.257 & 0.844 \\
\hline 1 & 0.124 & 0.117 & 0.087& 0.139 & 0.903 & 0.188 & 0.818 & 0.172 & 0.847 & 0.838 & 0.743 & 0.156 \\
\hline \text {Errors} & 0.035 & 0.033 & 0.028 & 0.033 & 0.026 & 0.037 & 0.033 & 0.038 & 0.033 & 0.035 & 0.030 & 0.037 \\
\hline
\end{array}
\nonumber
\end{equation}
\caption{Experimental data of phase estimation algorithm. The phase estimation of four unitary operations ($U=I$, $Z^{1/2}$, $Z^{5/4}$, and $Z^{3/2}$) correspond to phase shifts of $\varphi$ = 0, $\pi/2$, $5\pi/4$, $3\pi/2$, respectively. These phases can be accurately represented as $\varphi=2\pi \times (0.000,0.010,0.101,0.110)$ using 3-digit binary numbers. We estimate the phase of the corresponding $Z$-phase gate with three significant digits and obtain the probability that each digit is 0 or 1 by polarization measurement of the register. When the probability for 0 or 1 is greater than 1/2, this digit is selected. According to the data, we can estimate the phase of the four $Z$-phase gates with $100\%$ accuracy.} 
\label{table3}
\end{table}